\documentclass{aa}
\usepackage[varg]{txfonts}
\usepackage{natbib}  
\usepackage{graphicx}
\usepackage{color}
\usepackage{multirow}
\usepackage{mathtools}
\bibpunct{(}{)}{;}{a}{}{,}
\usepackage{tabularx}
\usepackage{amsmath}  
\usepackage{gensymb}
\usepackage{booktabs,siunitx}
\usepackage{dcolumn}

\begin{document} 

   \title{Interferometric view of the circumstellar envelopes of northern FU\,Orionis-type stars}
   \subtitle{} 
   \author{O. Feh\'er\inst{1,2}, \'A. K\'osp\'al\inst{1,3}, P. \'Abrah\'am\inst{1}, M. R. Hogerheijde\inst{4}, C. Brinch\inst{5}}
   \institute{Konkoly Observatory, Research Centre for Astronomy and Earth Sciences, Hungarian Academy of Sciences, H-1121 Budapest, Konkoly Thege Mikl\'os \'ut 15-17, Hungary
   \and
  	E\"otv\"os Lor\'and University, Department of Astronomy, P\'azm\'any P\'eter s\'et\'any 1/A, 1117 Budapest, Hungary
  	\and
  	Max-Planck-Institut für Astronomie, Königstuhl 17, D-69117 Heidelberg, Germany
   \and
	Leiden Observatory, Leiden University, Niels Bohrweg 2, 2333 CA, Leiden, The Netherlands
   \and
	Niels Bohr International Academy, The Niels Bohr Institute, University of Copenhagen, Blegdamsvej 17, 2100 Copenhagen \O, Denmark}
    
	\authorrunning{O. Feh\'er et al.}
	\titlerunning{Interferometric view of the envelopes of northern FUors}
   
%\date{Received <date> / Accepted <date>} 
 
  \abstract
   {FU\,Orionis-type objects are pre-main sequence, low-mass stars with large outbursts in visible light that last for several years or decades. They are thought to represent an evolutionary phase during the life of every young star when accretion from the circumstellar disk is enhanced during recurring time periods. These outbursts are able to rapidly build up the star while affecting the physical conditions inside the circumstellar disk and thus the ongoing or future planet formation. In many models infall from a circumstellar envelope seems to be necessary to trigger the outbursts.}
   {We characterize the morphology and the physical parameters of the circumstellar material around FU\,Orionis-type stars using the emission of millimeter wavelength molecular tracers. The high spatial resolution study gives insight into the evolutionary state of the objects, the distribution of parameters in the envelopes and the physical processes forming the environment of these stars.}
   {We observed the J=1$-$0 rotational transition of $^{13}$CO and C$^{18}$O towards eight northern FU\,Orionis-type stars (V1057\,Cyg, V1515\,Cyg, V2492\,Cyg, V2493\,Cyg, V1735\,Cyg, V733\,Cep, RNO\,1B and RNO\,1C) and determine the spatial and velocity structure of the circumstellar gas on the scale of a few thousands of AU. We derive temperatures and envelope masses and discuss the kinematics of the circumstellar material.}
   {We detected extended CO emission associated with all our targets. Smaller scale CO clumps were found to be associated with five objects with radii of 2000$-$5000\,AU and masses of 0.02$-$0.5\,$M_{\odot}$; these are clearly heated by the central stars. Three of these envelopes are also strongly detected in the 2.7\,mm continuum. No central CO clumps were detected around V733\,Cep and V710\,Cas that can be interpreted as envelopes but there are many other clumps in their environments. Traces of outflow activity were observed towards V1735\,Cyg, V733\,Cep and V710\,Cas.}
   {The diversity of the observed envelopes enables us to set up an evolutionary sequence between the objects. We find their evolutionary state to range from early, embedded Class\,I stage to late, Class\,II-type objects with very low-mass circumstellar material. We also find evidence of larger scale circumstellar material influencing the detected spectral features in the environment of our targets. These results reinforce the idea of FU\,Orionis-type stars as representatives of a transitory stage between embedded Class\,I young stellar objects and classical T-Tauri stars.}

   \keywords{molecular data - stars: pre-main sequence - stars: variables: T Tauri, Herbig Ae/Be - (stars:) circumstellar matter - ISM: structures}
   \maketitle
%-------------------------------------------------------------------

\section{Introduction}

\begin{table*}[ht]
	\centering
    \footnotesize
		\caption{Observed targets and log of the observations}
       % \resizebox{\textwidth}{!}{
		\begin{tabular}{l l l l l c c c c c}
		\hline
		\multicolumn{1}{c}{Target} & other IDs & \multicolumn{1}{c}{RA (J2000)} & \multicolumn{1}{c}{DEC (J2000)} & \multicolumn{1}{c}{D} & \multicolumn{1}{c}{Telescope}  & \multicolumn{1}{c}{Date} & \multicolumn{1}{c}{Beam} & \multicolumn{1}{c}{Merged beam} & \multicolumn{1}{c}{PA}  \\
        \multicolumn{1}{c}{\ } &  & \multicolumn{1}{c}{[h:m:s]} & \multicolumn{1}{c}{[$^{\circ}$:$'$:$''$]} & \multicolumn{1}{c}{[pc]} & \multicolumn{1}{c}{\ }  & \multicolumn{1}{c}{\ } & \multicolumn{1}{c}{[arcsec]} & \multicolumn{1}{c}{[arcsec]} &  \multicolumn{1}{c}{[deg]} \\
		\hline 
		\hline
        V1057 Cyg & HBC 300 & 20:58:53.73 & 44:15:28.5 & 600 & PdBI & 28 March, 2 April 2012 & 2.7\,$\times$\,2.2 & \multirow{2}{*}{2.8\,$\times$\,2.3} & \multirow{2}{*}{95}\\
	&	&  &  &	& 30m & 19-22 June, 2012 & 23.5 &  \\ 
V1515\,Cyg & HBC 692 & 20:23:48.02 & 42:12:25.8 & 1000 & PdBI & 28 March, 2 April 2012 & 2.7\,$\times$\,2.2 & \multirow{2}{*}{2.8\,$\times$\,2.3}  & \multirow{2}{*}{95}\\
	& 	&  & 	&  & 30m & 20-22 June, 2012 & 23.5 & \\
V2492\,Cyg & PTF 10nvg  & 20:51:26.23 & 44:05:23.9 & 550 &PdBI & 28 March, 2 April 2012 & 2.8\,$\times$\,2.2 & \multirow{2}{*}{2.8\,$\times$\,2.3} & \multirow{2}{*}{95}\\
	& 	& 	& &  & 30m & 20-22 June, 2012 & 23.5 & \\
V2493\,Cyg & HBC 722 & 20:58:17.03 & 43:53:43.4 & 550 & PdBI & 28 March, 2 April 2012 & 2.7\,$\times$\,2.2 & \multirow{2}{*}{2.8\,$\times$\,2.3}   & \multirow{2}{*}{95}\\
	& 	& 	& &  & 30m & 20-22 June, 2012 & 23.5 & \\
V1735\,Cyg & Elias 1-12 & 21:47:20.66 & 47:32:03.6 & 850 & PdBI & 5 Apr 2014 & 2.4\,$\times$\,2.2 & \multirow{2}{*}{2.4\,$\times$\,2.2}    & \multirow{2}{*}{29.8}\\
	& 	&	& &  & 30m & 25 June 2014 & 23.6 &  \\
V733\,Cep & Persson's Star & 22:53:33.26 & 62:32:23.6 & 900 & PdBI & 4, 18 Apr 2014 & 2.4\,$\times$\,2.3 & \multirow{2}{*}{2.5\,$\times$\,2.3}    & \multirow{2}{*}{43.6}\\
	&	&	& &  & 30m &	26 June 2014 & 23.6 & \\
V710\,Cas & RNO 1B/C & 00:36:46.30 & 63:28:54.1 & 800 & PdBI & 6-7,17 Apr 2014 & 2.5\,$\times$\,2.2 & \multirow{2}{*}{2.5\,$\times$\,2.4}   & \multirow{2}{*}{69.6}\\
	&	&	& & & 30m & 27 June 2014 & 23.6 & \\
\hline
		\end{tabular}%}
    \tablefoot{The columns are: (1,2) target name and alternative name; (3,4) equatorial coordinates of the optical position of the target; (5) distance; (6) telescope; (7) observing dates; (8) single-dish or synthetic beam HPBW;  (9) synthetic beam HPBW of the merged observations; (10) position angle of the synthetic beam of the merged observations.}
	\label{obslog}
\end{table*}

The class of low-mass pre-main sequence objects that show sudden, 5$-$6 magnitude brightness increase in the optical and near-infrared (NIR) regime was first described by \citet{herbig1977}. These FU\,Orionis-type stars (FUors) remain bright for years or decades, sometimes fading slowly. The cause of the outburst is attributed to enhanced accretion (from a typical rate of 10$^{-7}$ to 10$^{-4}$\,M$_{\odot}$yr$^{-1}$) from the circumstellar disk to the surface of the star \citep{hartmann1996}. During one outburst up to 10$^{-2}$\,M$_{\odot}$ mass can be accreted to the central star, playing a large role in accumulating the final stellar mass. The episodic FUor outbursts are predicted to change the parameters and structure of the circumstellar disk and envelope, e.g. structural changes of the disk \citep{mosoni2013}, signs of evaporation \citep{kun2011} or crystallization \citep{abraham2009}, the opening of the cavity \citep{ruiz2017} and the movement of the snowline \citep{cieza2016}. Theoretical explanations for the outburst phenomenon include thermal instabilities \citep{bell1994}, gravitational instabilities \citep{armitage2001, boley2006} and the perturbation of a close stellar or sub-stellar companion \citep{bonnell1992}. Nevertheless, continuous infall from a circumstellar envelope is needed to replenish the accreted material in the disk and certain studies show that this process might also play a role in triggering the outbursts \citep{vorobyov2006, vorobyov2013}. 

In the optical regime FUors show similar spectra, with F/G supergiant spectral types and broad absorption lines, while in the NIR they show K/M supergiant spectral types. The optical metallic lines (Fe I, Li I, Ca I) and the infrared CO absorption line are double-peaked and broadened, which is consistent with a rotating disk origin. On the contrary, in the mid-infrared (MIR) they exhibit much higher variations, showing strong or weak silicate emission and even absorption. The far-infrared (FIR) and sub-millimeter continuum may also appear weaker or stronger in the different sources \citep[][and references therein]{audard2014}. This diversity of the observed MIR and FIR FUor spectra is explained by \citet{quanz2007a} proposing an evolutionary sequence that span from younger FUors that are still embedded into dense, dusty envelopes, resembling Class\,I objects to more evolved stars showing many properties of Class\,II objects with a smaller, remnant envelope. Theoretical models \citep{vorobyov2015} predict that after a sequence of eruptions, the stars enter a more quiescent phase, becoming classical T Tauri stars. Thus, FUors might represent the link between Class\,I and Class\,II low-mass young stars. The outbursts are often accompanied by jets \citep{whelan2010}, Herbig-Haro objects \citep{reipurth1983, bally2003}, molecular outflows \citep{evans1994} and reflection nebulae \citep{goodrich1987, quanz2007a}.

Millimeter wavelength observations of molecular line and dust continuum emission provide information about the small-scale structure and kinematics of the disk and the envelope around FUors. Low and high-density circumstellar material can be mapped with transitions of $^{12}$CO, $^{13}$CO and C$^{18}$O \citep{kospal2011, hillenbrand2013, kospal2016, kospal2017}, the outflows with $^{12}$CO \citep{evans1994}, the cavity walls with HCO$^{+}$ and HCN emission, while SiO and SO emission traces the shocked material inside the outflow \citep{hogerheijde1999}. There are relatively few high spatial resolution studies of FUors \citep{momose1998, hales2015, kospal2016, zurlo2017, ruiz2017}. 

Here we present a $^{13}$CO(1$-$0) and C$^{18}$O(1$-$0) line survey with the aim of investigating the spatial and kinematic structure of the gas around eight FU\,Orionis-type young stars with a high spatial and spectral resolution interferometric survey. We calculate optical depths, excitation temperatures, column densities and continuum masses around the targets in Sect. \ref{sectres}, we describe the environment of the stars in detail and calculate line-based masses in Sect. \ref{sectanalysis}, and we identify the envelopes, compare our study with previous results and assess the evolutionary state of the observed FUors in Sect. \ref{sectdisc}.

\section{Observations and data reduction}

We observed seven northern FUor systems with the Plateau de Bure Interferometer (PdBI) and the IRAM 30\,m telescope. The field centered on V710\,Cas (RNO\,1B) contains another close-by FUor, RNO\,1C, with a separation of 6$\arcsec$. Table \ref{obslog} lists the targets and the details of the observations. 

\object{V1057 Cyg}, \object{V1515\,Cyg}, \object{V2492\,Cyg} and \object{V2493\,Cyg} were observed with the PdBI on 28 March and on 2 April, 2012. The antennas were in the 6Cq configuration, providing \textit{uv} coverage between 15\,m and 175\,m. The observations were done on the first night when the sources were setting and on the second night when they were rising, and each target was measured for about 10\,minutes, before switching to the next one, thus providing the best possible \textit{uv} coverage. The total on-source correlation time for each target was 2 hours. We used the 3\,mm receiver centered on 109.0918\,GHz, halfway between the $^{13}$CO(1-0) and C$^{18}$O(1-0) lines and each line was measured with a 20\,MHz wide correlator with 39\,kHz resolution. We also used two 160\,MHz wide correlators to measure the 2.7\,mm continuum emission. The single dish half-power beam width (HPBW) at this wavelength is 45.8$\arcsec$. Bright quasars were used for radio frequency bandpass, phase, and amplitude calibration and the flux scale was determined using the carbon star \object{MWC 349}. The weather conditions were stable throughout the observations, with precipitable water vapor between 5$-$9\,mm. The rms (root mean square) phase noise was typically below 30$^{\circ}$ and the flux calibration accuracy is estimated to be around 15\%. The single-dish observations were performed with the 30\,m telescope on 3 nights between 20$-$22 June, 2012. We obtained Nyquist-sampled 112$\arcsec\times$\,112$\arcsec$ on-the-fly maps with 7$\arcsec$ spacing using the Eight Mixer Receiver (EMIR) in frequency switching mode with 3.9\,MHz frequency throw in the 110\,GHz band with the Versatile Spectrometer Array (VESPA) that provided 20\,MHz bandwidth with 19\,kHz channel spacing. The $^{13}$CO and C$^{18}$O lines were covered in the UI sideband in this setup. The weather conditions were good, with precipitable water vapor between 2$-$8\,mm on 28\,March and between 2$-$4\,mm on 2 April. \object{V1735\,Cyg}, \object{V733\,Cep} and \object{V710\,Cas} were observed with the PdBI between 4$-$7 April and 17$-$18 April, 2014 and with the 30\,m telescope between 25$-$27 June, 2014. Both the interferometric and the single-dish observations were made with the same instruments and settings as during the 2012 observing run. The weather conditions were good during both the single dish and the interferometric observing sessions with precipitable water vapor between 1$-$15\,mm.

\begin{table*}[htp]
	\centering
   % \footnotesize
		\caption{Parameters of the detected continuum sources.}
        \resizebox{\textwidth}{!}{
		\begin{tabular}{l l l l l l l l l l l l l l l l l l l l}
		\hline
		\multicolumn{1}{c}{Target} & \multicolumn{1}{c}{RA} & \multicolumn{1}{c}{DEC} & \multicolumn{1}{c}{a$_{\rm conv}$} & \multicolumn{1}{c}{b$_{\rm conv}$} & \multicolumn{1}{c}{PA$_{\rm conv}$}  & \multicolumn{1}{c}{a$_{\rm deconv}$} & \multicolumn{1}{c}{b$_{\rm deconv}$} & \multicolumn{1}{c}{PA$_{\rm deconv}$} & \multicolumn{1}{c}{$S_{\nu}$}  & \multicolumn{1}{c}{S$\mathrm{_{\nu,p}}$} & $M_{\rm cont}$  \\
		\multicolumn{1}{c}{\,} & \multicolumn{1}{c}{[h:m:s]} & \multicolumn{1}{c}{[$^{\circ}$:$'$:$''$]} & \multicolumn{1}{c}{[arcsec]} & \multicolumn{1}{c}{[arcsec]} & \multicolumn{1}{c}{[deg]}  & \multicolumn{1}{c}{[arcsec]} & \multicolumn{1}{c}{[arcsec]} & \multicolumn{1}{c}{[deg]} & \multicolumn{1}{c}{[mJy]}  & \multicolumn{1}{c}{[mJybeam$^{-1}$]} & [$M_{\odot}$]\\
		\hline 
		\hline
V1057 Cyg & 20:58:53.72 & 44:15:28.01 & 3.4 $\pm$ 0.1 & 3.2 $\pm$ 0.1 & 176 $\pm$ 37 & 2.6 $\pm$ 0.2 & 1.8 $\pm$ 0.3 &   4 $\pm$ 36 & 4.9 $\pm$ 0.2 & 2.62 $\pm$ 0.04 & 0.39 \\
V1515\,Cyg  & 20:23:47.95 & 42:12:25.80 & ... & ... & ... & ... & ... & ... & ... & 0.18 $\pm$ 0.03 & 0.04 \\
V2492\,Cyg  & 20:51:26.21 & 44:05:23.83 & 4.3 $\pm$ 0.2 & 2.7 $\pm$ 0.1 &  85 $\pm$  2 & 3.3 $\pm$ 0.2 & 1.5 $\pm$ 0.2 &  82 $\pm$  3 & 5.0 $\pm$ 0.2 & 2.65 $\pm$ 0.04 & 0.34 \\
V2493\,Cyg  & ... & ... & ... & ... & ... & ... & ... & ... & ... &  <\,0.24 & <\,0.02\\
\multicolumn{1}{r}{MMS1} & 20:58:16.52 & 43:53:53.34 & 3.0 $\pm$ 0.1 & 2.5 $\pm$ 0.1 &  98 $\pm$  5 & 1.3 $\pm$ 0.2 & 1.1 $\pm$ 0.3 & 115 $\pm$ 81 & 4.5 $\pm$ 0.2 & 3.63 $\pm$ 0.08 \\
\multicolumn{1}{r}{MMS2} & 20:58:16.79 & 43:53:35.61 & 2.9 $\pm$ 0.2 & 2.3 $\pm$ 0.1 &  99 $\pm$  9 & 1.1 $\pm$ 0.6 & 0.6 $\pm$ 0.4 & 112 $\pm$ 33 & 1.6 $\pm$ 0.2 & 1.44 $\pm$ 0.08  \\
\multicolumn{1}{r}{MMS3} & 20:58:17.66 & 43:53:30.20 & 5.6 $\pm$ 0.5 & 3.2 $\pm$ 0.2 &  51 $\pm$  4 & 5.0 $\pm$ 0.5 & 2.0 $\pm$ 0.4 &  47 $\pm$  5 & 4.1 $\pm$ 0.4 & 1.37 $\pm$ 0.08 \\
\multicolumn{1}{r}{MMS4} & 20:58:16.33 & 43:53:25.97 & 4.3 $\pm$ 0.4 & 3.1 $\pm$ 0.2 &  41 $\pm$  9 & 3.6 $\pm$ 0.6 & 1.6 $\pm$ 0.7 &  34 $\pm$ 12 & 4.4 $\pm$ 0.5 & 2.02 $\pm$ 0.08  \\
 V1735\,Cyg & 21:47:20.65 & 47:32:03.77 & 2.9 $\pm$ 0.2 & 2.3 $\pm$ 0.1 &  19 $\pm$  8 & 1.6 $\pm$ 0.3 & 0.9 $\pm$ 0.4 &  15 $\pm$ 22 & 2.3 $\pm$ 0.2 & 1.79 $\pm$ 0.05 & 0.37 \\
V733\,Cep  & 22:53:33.47 & 62:32:23.82 & ... & ... & ... & ... & ... & ... & ... & 0.38 $\pm$ 0.10 & 0.07 \\
V710\,Cas & ... & ... & ... & ... & ... & ... & ... & ... & ... & <\,0.36 & <\,0.05 \\
  \multicolumn{1}{r}{GM\,1-33} & 00:36:47.36 & 63:29:02.33 & 2.6 $\pm$ 0.2 & 2.2 $\pm$ 0.2 &  14 $\pm$ 18 & <\,1.6 & <\,0.9 &  ...  & 4.4 $\pm$ 0.6 & 4.14 $\pm$ 0.12 \\
\multicolumn{1}{r}{RNO\,1D} & 00:36:46.66 & 63:28:56.94 & 4.8 $\pm$ 0.6 & 4.6 $\pm$ 0.5 &  21 $\pm$ 82 & 4.2 $\pm$ 0.8 & 4.0 $\pm$ 0.8 &  39 $\pm$ 82 & 9.4 $\pm$ 1.3 & 2.27 $\pm$ 0.12  \\
 & 00:36:46.70 & 63:28:49.47 & 4.1 $\pm$ 0.4 & 3.9 $\pm$ 0.4 & 118 $\pm$ 78 & 3.5 $\pm$ 0.7 & 3.0 $\pm$ 0.6 & 106 $\pm$ 86 & 9.0 $\pm$ 1.2 & 2.97 $\pm$ 0.12 \\
\hline
		\end{tabular}}
    \tablefoot{The columns are: (1) target name; (2,3) equatorial coordinates of the continuum source; (4,5,6) major, minor axis and position angle of the fitted 2D Gaussian (convolved with beam); (7,8,9) major axis, minor axis, and position angle of fitted 2D Gaussian (deconvolved); (10) integrated flux; (11) peak flux; (12) continuum mass.}
	\label{continuum_data}
\end{table*}

\begin{figure*}
\includegraphics[width=\linewidth, trim=0cm 0cm 0cm 0cm]{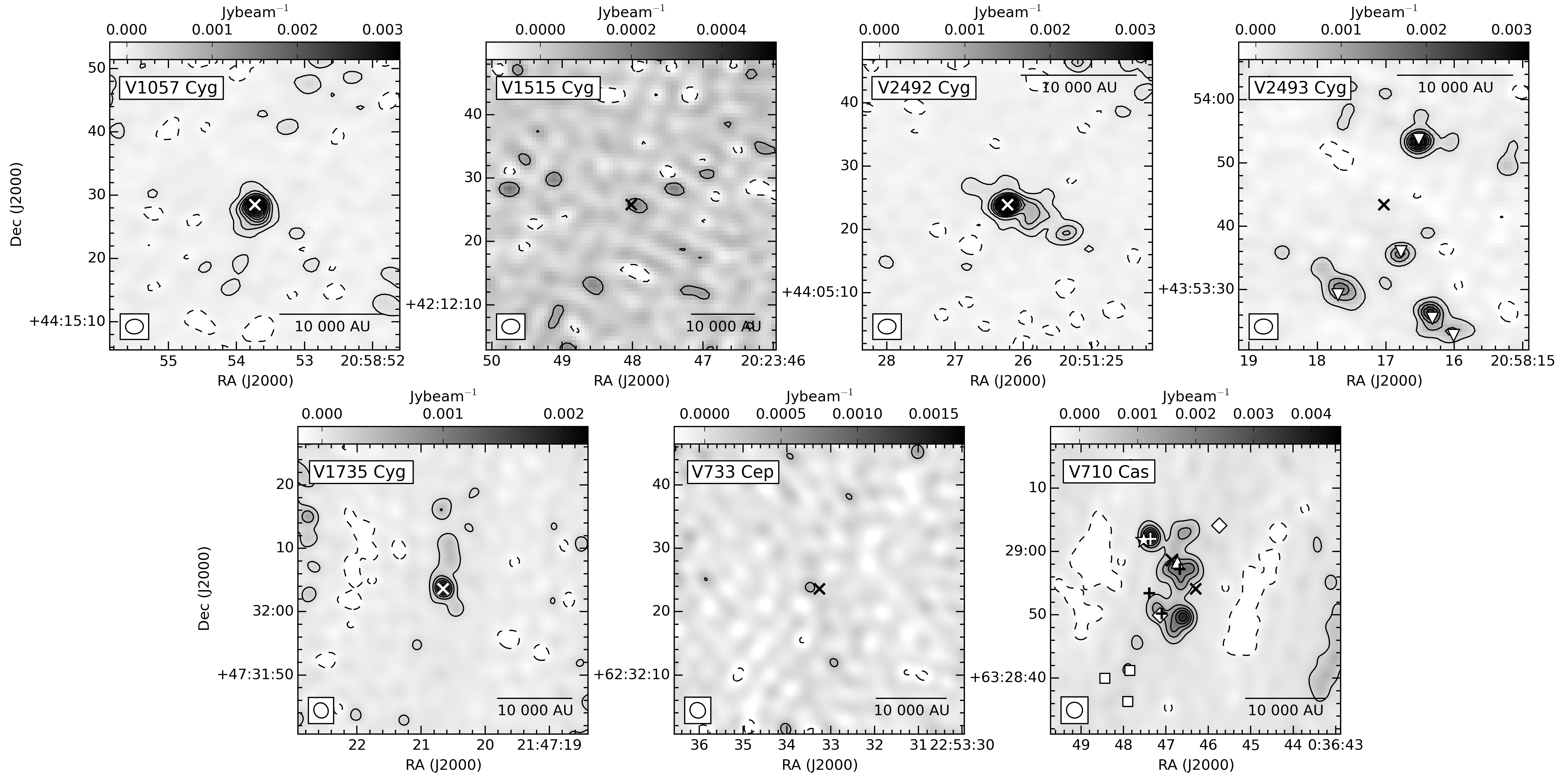}
\caption{2.7\,mm continuum maps of the sources. No short-spacing data was combined with the interferometric observations of the continuum. The solid lines mark the 3$\sigma$, 9$\sigma$, 15$\sigma$... contour levels and dashed lines show the -3$\sigma$ contour level. Crosses mark the positions of the FUors and the beam is shown in the bottom left corner. Other associated sources are marked as follows: upside down triangles mark the millimeter sources around V2493\,Cyg from \citet{dunham2012}, in the region around V710\,Cas (RNO\,1B) a second cross marks RNO\,1C, an asterisk marks IRAS\,00338+6312 \citep{staude1991}, crosses mark the radio sources from \citet{anglada1994}, triangle marks the sub-millimeter source from \citet{sandell2001}, squares mark the IRAC sources from \citet{quanz2007b} and diamonds mark RNO\,1F and RNO\,1G.}
\label{continuum}
\end{figure*}

\begin{figure*}[ht]
\centering
\includegraphics[angle=-90,width=.245\linewidth, trim=1cm 4cm 2cm 1cm]{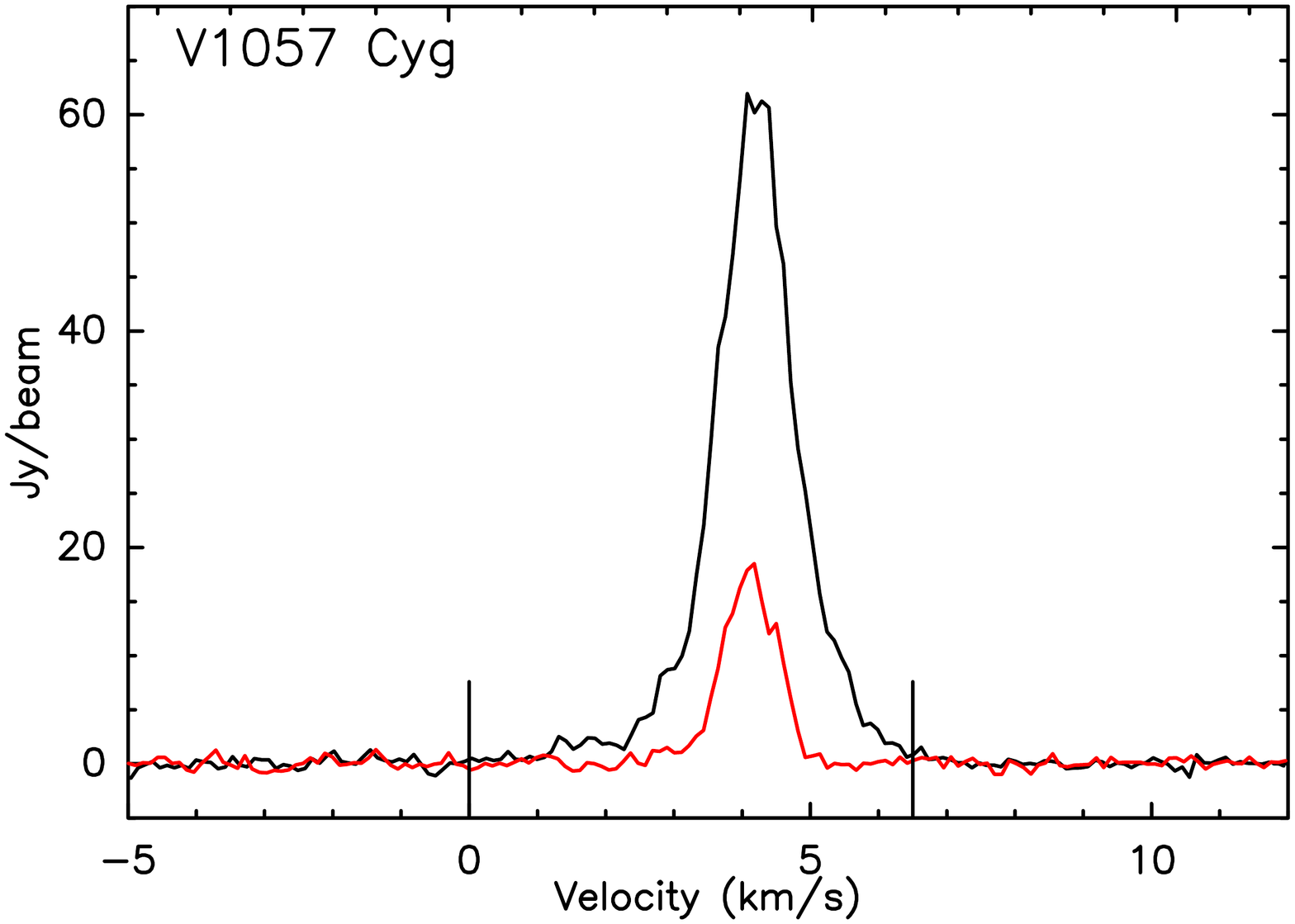}
\includegraphics[angle=-90,width=.245\linewidth, trim=1cm 4cm 2cm 1cm]{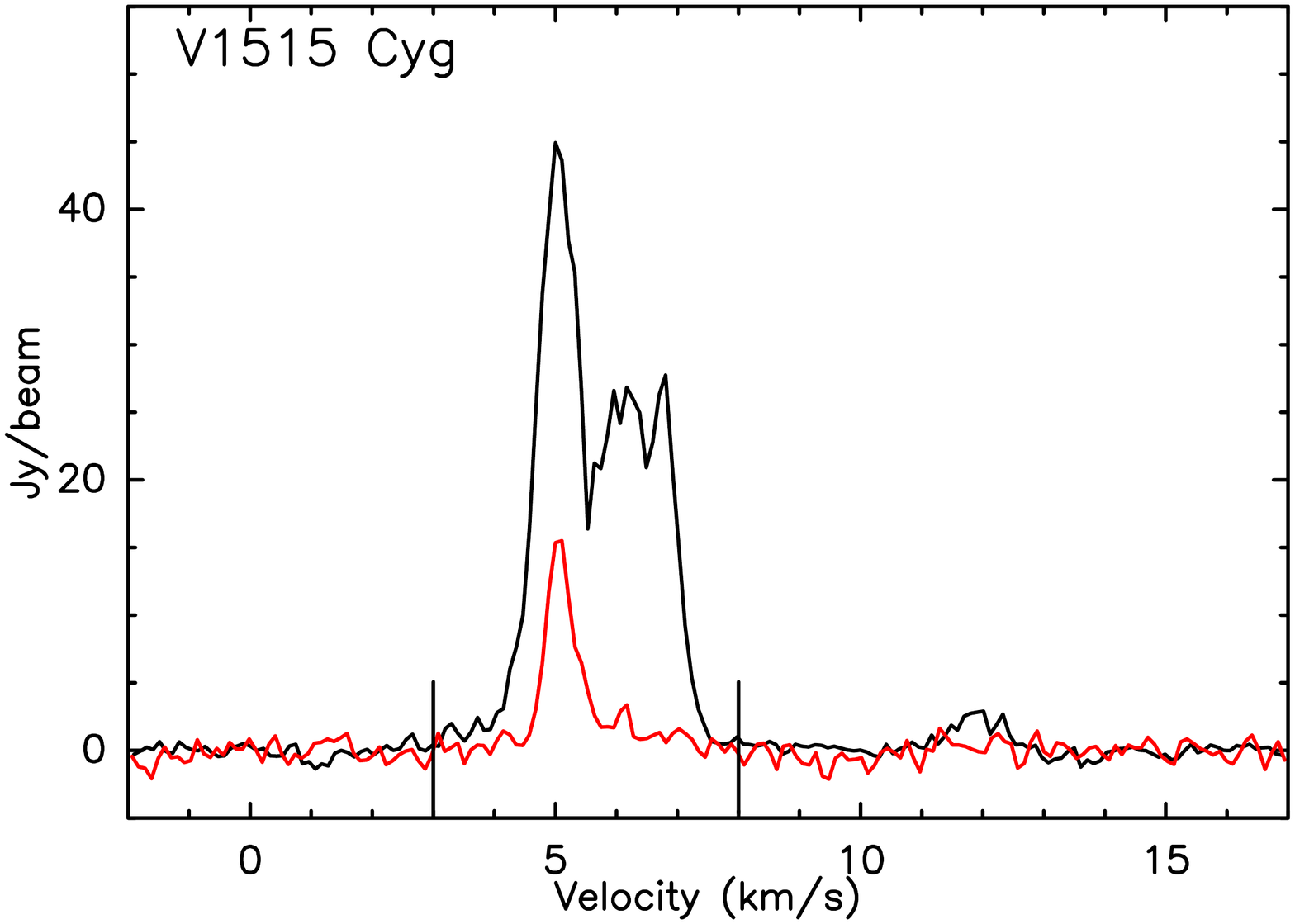} 
\includegraphics[angle=-90,width=.245\linewidth, trim=1cm 4cm 2cm 1cm]{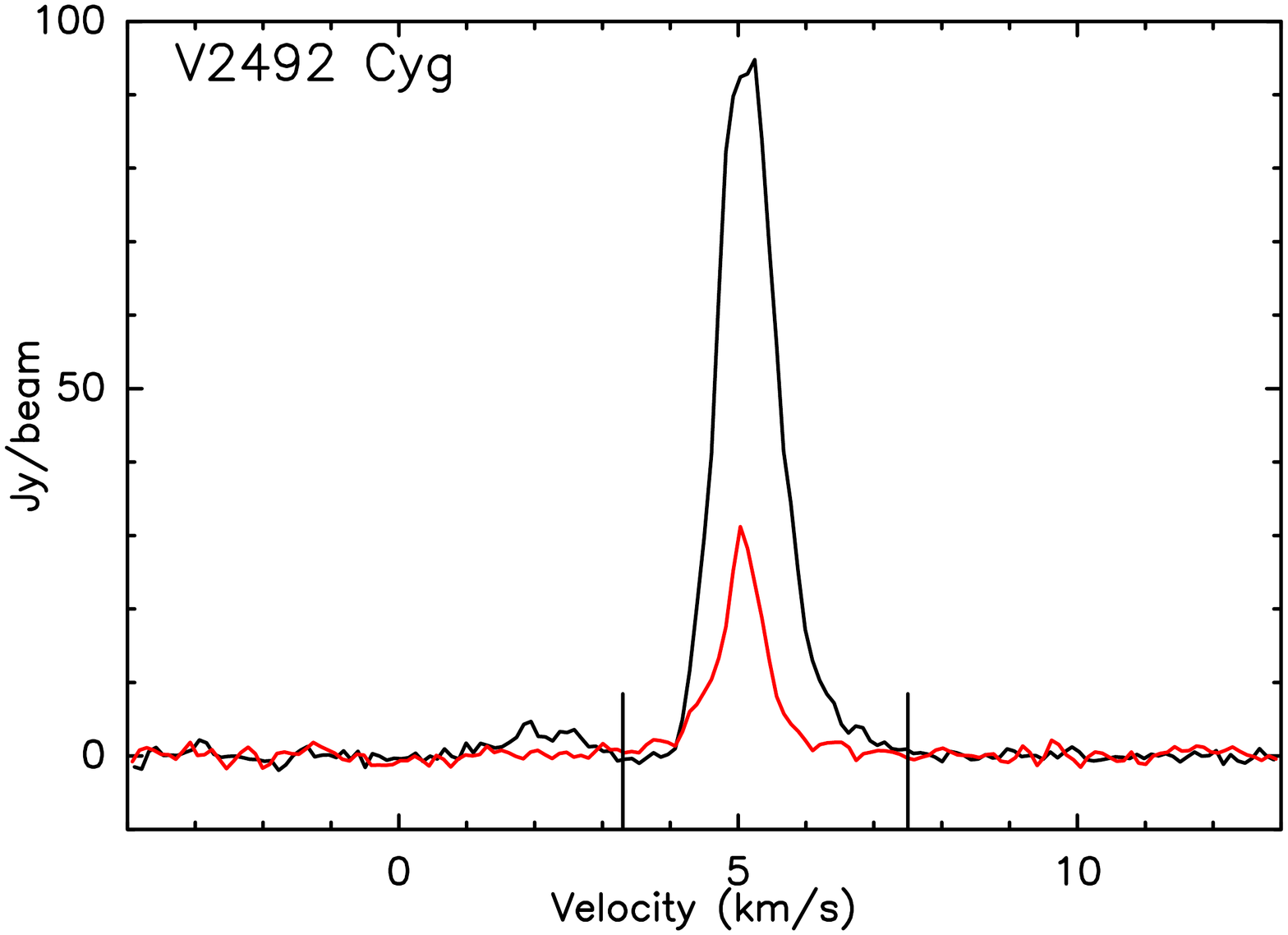}
\includegraphics[angle=-90,width=.245\linewidth, trim=1cm 4cm 2cm 1cm]{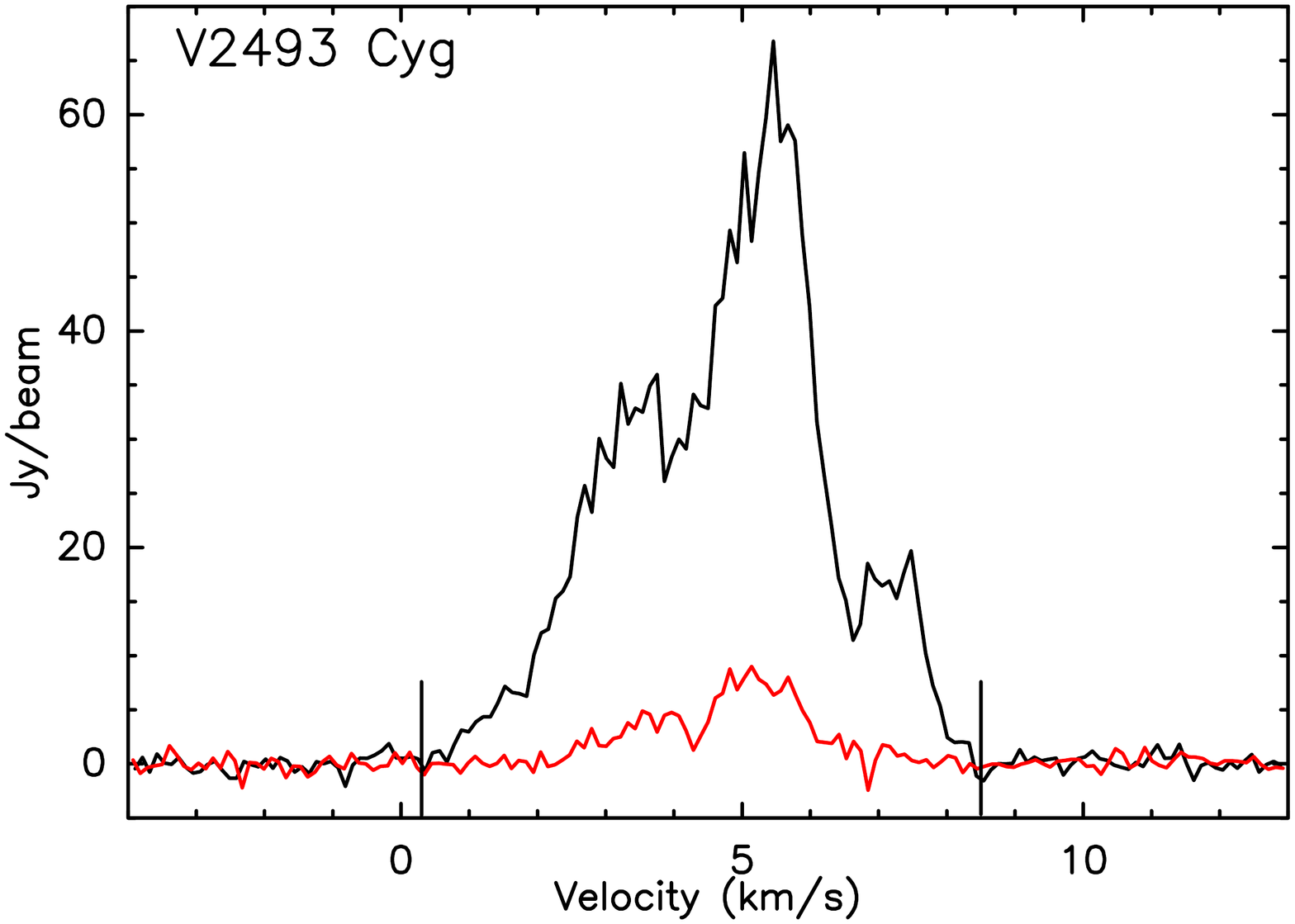}\\
\centering
\includegraphics[angle=-90,width=.245\linewidth, trim=1cm 4cm 2cm 1cm]{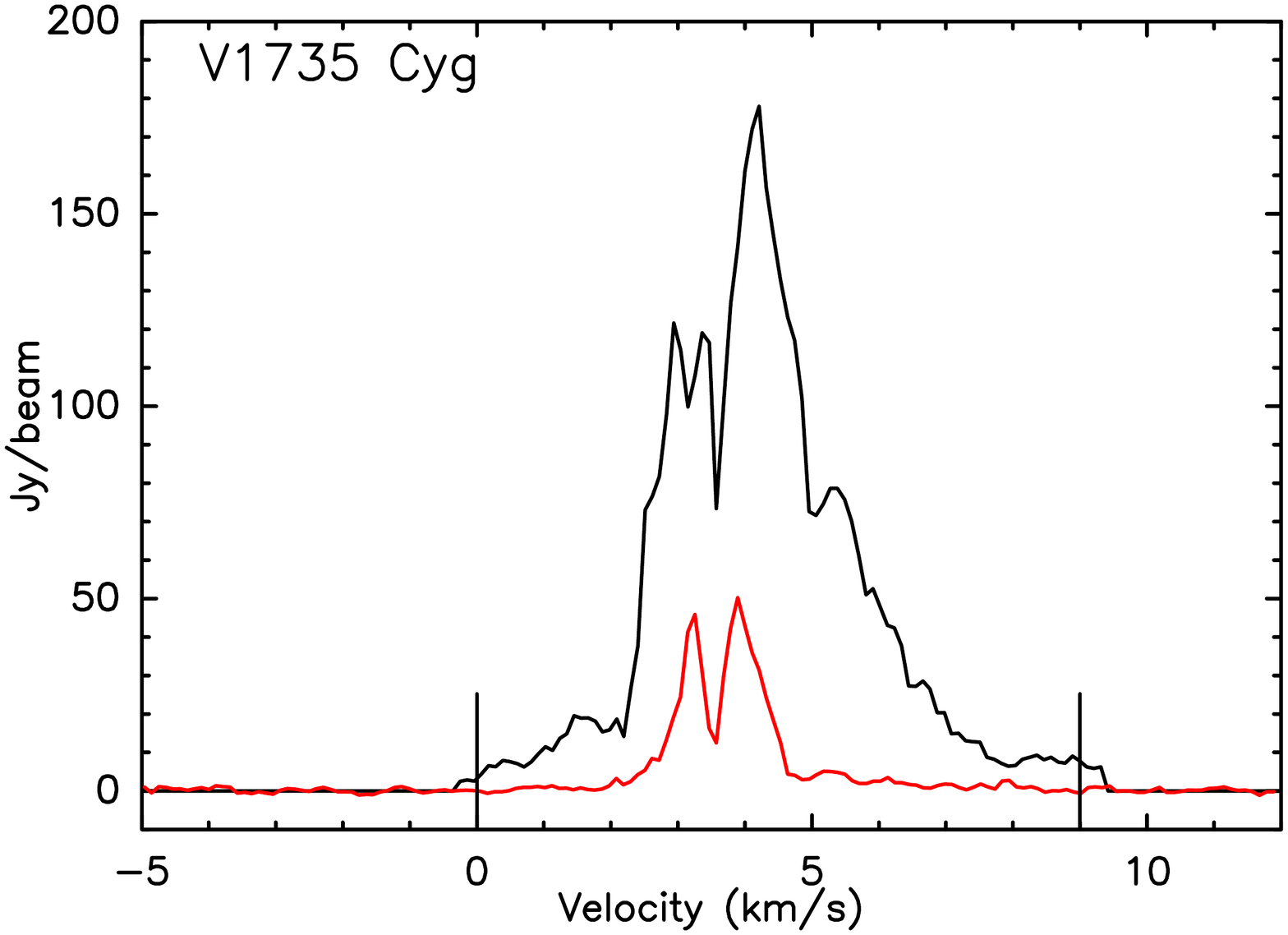}
\includegraphics[angle=-90,width=.245\linewidth, trim=1cm 4cm 2cm 1cm]{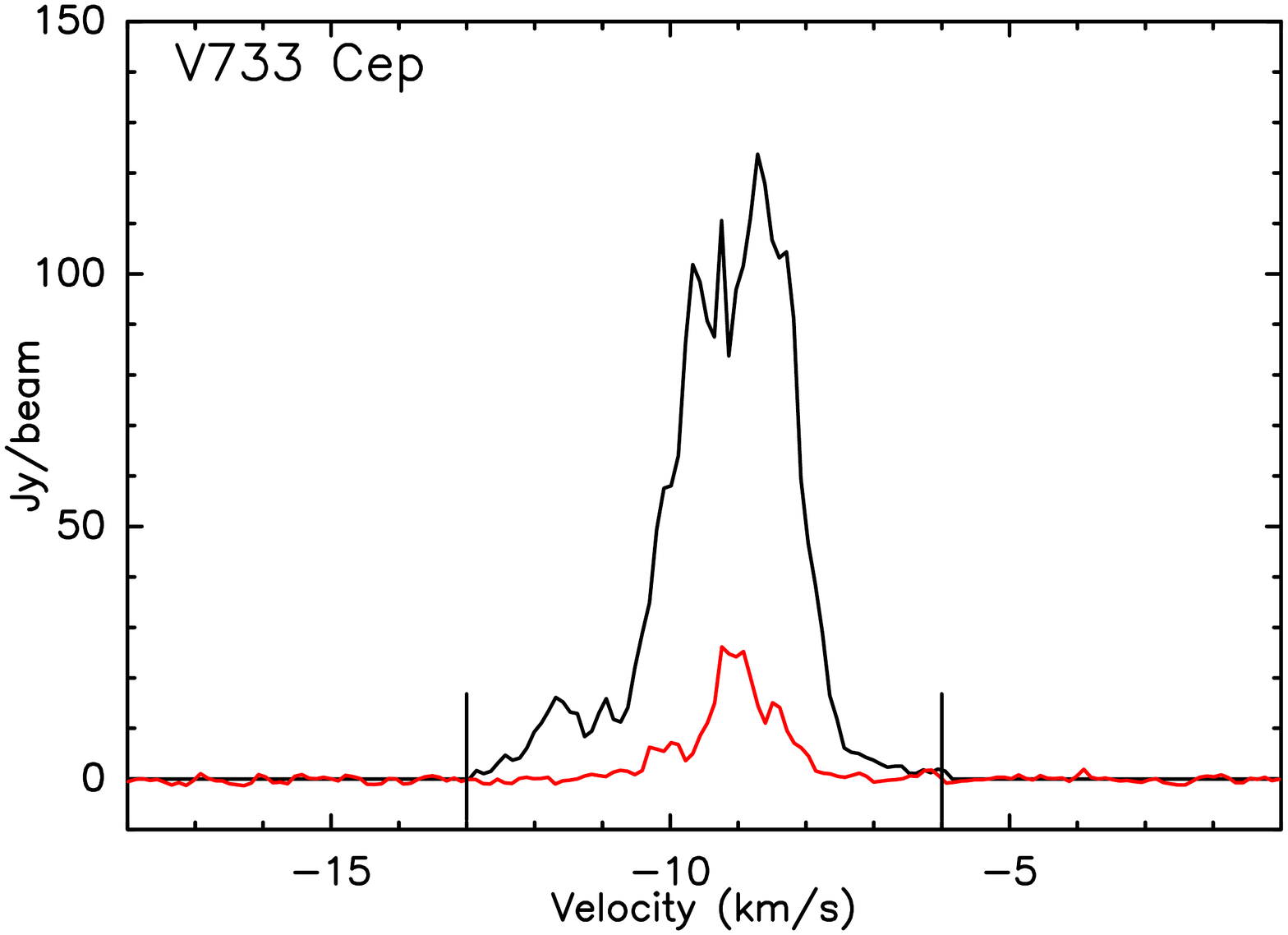}
\includegraphics[angle=-90,width=.245\linewidth, trim=1cm 4cm 2cm 1cm]{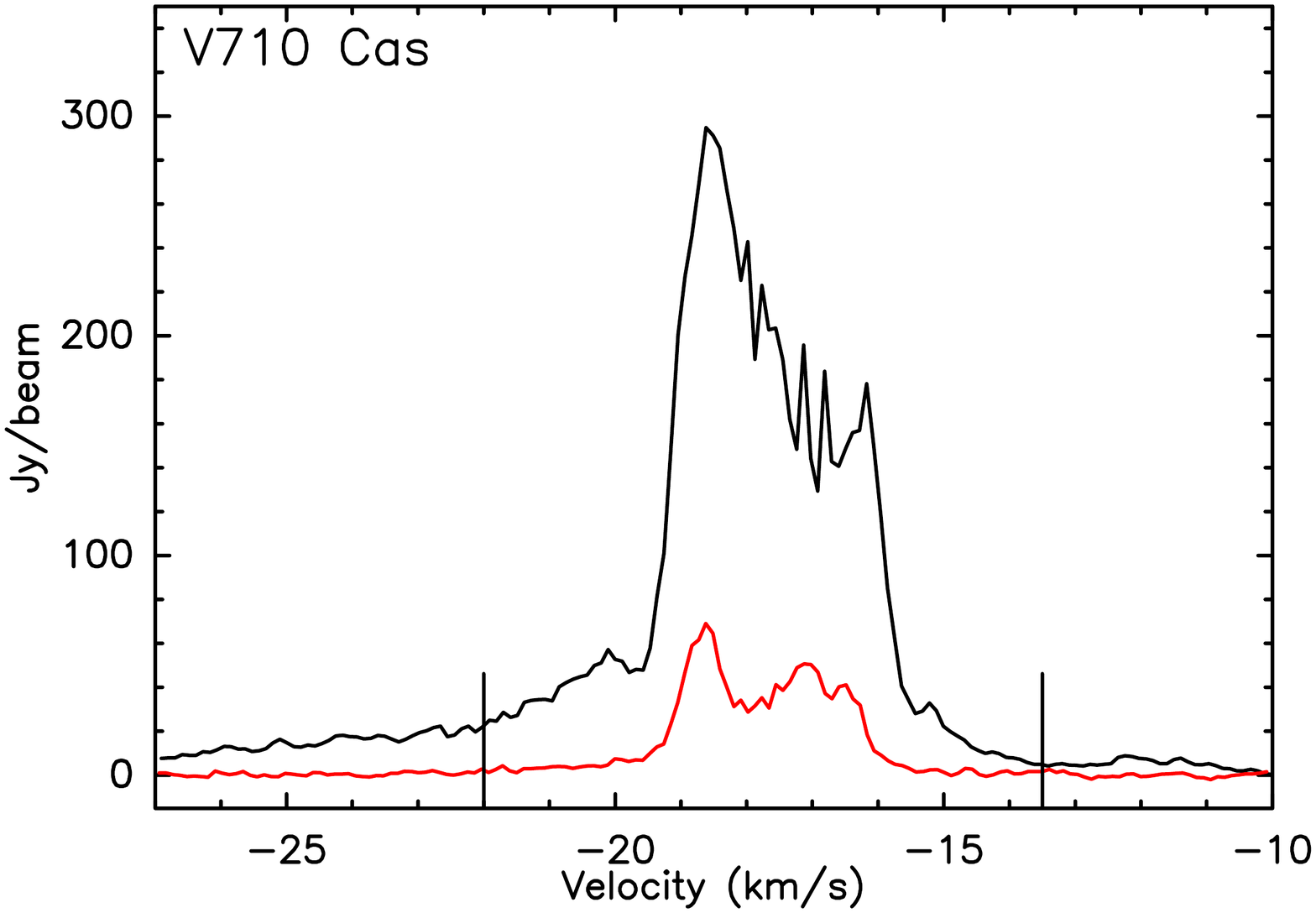}
	\caption{$^{13}$CO(1$-$0) (black) and C$^{18}$O(1$-$0) (red) spectra integrated over the beam size ($\approx$\,2.5$\arcsec$) around the center position of our maps. Vertical lines mark the velocity range where the moments shown in Figs.\ref{v1057_fig}-\ref{v710_fig} were derived.}
	\label{avespec}
\end{figure*} 

\newcolumntype{L}{D{.}{.}{2,2}}
\begin{table*}[tp]
	\centering
	\sisetup{separate-uncertainty=true}
   % \footnotesize
		\caption{Parameters of the spectra at the center of each map averaged over one beam.}
        \resizebox{.75\textwidth}{!}{
		\begin{tabular}{l c c c c c r}
		\hline
		\multicolumn{1}{c}{Target} & \multicolumn{1}{c}{$v\mathrm{_{LSR}}$($^{13}$CO)} & \multicolumn{1}{c}{$\Delta v$($^{13}$CO)} & \multicolumn{1}{c}{S$\mathrm{_{p}}$($^{13}$CO)}  & \multicolumn{1}{c}{$v\mathrm{_{LSR}}$(C$^{18}$O)} & \multicolumn{1}{c}{$\Delta v$(C$^{18}$O)} & \multicolumn{1}{c}{S$\mathrm{_{p}}$(C$^{18}$O)}  \\
\multicolumn{1}{c}{ \,} & \multicolumn{1}{c}{[kms$^{-1}$]} & \multicolumn{1}{c}{[kms$^{-1}$]} & \multicolumn{1}{c}{[mJybeam$^{-1}$]}  &  \multicolumn{1}{c}{[kms$^{-1}$]} & \multicolumn{1}{c}{[kms$^{-1}$]} & \multicolumn{1}{c}{[mJybeam$^{-1}$]}  \\
		\hline 
		\hline
		\midrule
V1057 Cyg &   4.05 & 1.97 &   400 $\pm$     1 &   4.10 & 1.18 &    79 $\pm$     1 \\
V1515\,Cyg &   5.80 & 1.97 &   397 $\pm$     2 &   5.38 & 1.66 &   147 $\pm$     2 \\
V2492\,Cyg &   4.97 & 1.17 &   480 $\pm$     2 &   4.90 & 1.00 &   117 $\pm$     1 \\
V2493\,Cyg &   4.65 & 3.44 &   507 $\pm$     1 &   4.79 & 2.50 &    75 $\pm$     1 \\
V1735\,Cyg &   4.16 & 3.87 &   418 $\pm$     1 &   3.89 & 2.63 &   139 $\pm$     2 \\
V733\,Cep &  $-$9.23 & 2.50 &   424 $\pm$     2 &  $-$8.99 & 1.65 &    97 $\pm$     2 \\
V710\,Cas & $-$17.83 & 3.38 &   943 $\pm$     8 & $-$17.72 & 2.87 &   163 $\pm$     2 \\
\hline
		\end{tabular}}
    \tablefoot{The columns are: (1) target name; (2) intensity averaged velocity of the $^{13}$CO line (first order moment); (3) $^{13}$CO linewidth (from the second order moment); (4) peak flux density of the $^{13}$CO line; (5)-(7) the same parameters for the C$^{18}$O line.}
	\label{avespec_data}
\end{table*}

The data reduction of the single-dish spectra was done with the GILDAS-based CLASS and GREG\footnote{http://www.iram.fr/IRAMFR/GILDAS} software packages. We identified lines in the folded spectra, discarded those parts of the spectra where the negative signal from the frequency switching appeared, and subtracted a 4th order polynomial baseline. The interferometric observations were reduced in the standard way with the GILDAS-based CLIC application, resulting in a primary beam of 45.8$\arcsec$ and synthetic beams close to 2.5\,$\arcsec$. For the line spectroscopy we merged the single-dish and interferometric measurements in the $uv$-space to correctly recover both the smaller and larger scale structure of the sources. First we took the Fourier transform of the single-dish images, then determined the shortest \textit{uv}-distance in the interferometric dataset. A regular grid was made that filled out a circle within this shortest \textit{uv}-distance, then we took the data points at these locations from the Fourier-transformed single-dish image and merged them with the original interferometric dataset. For the continuum only the PdBI observations were combined. After this step the imaging and cleaning was done in the usual manner. 

\section{Results}
\label{sectres}

\subsection{Methods of CO line analysis}

The optical depth of the two isotopologues, $\tau_{13}$ and $\tau_{18}$ can be determined from the ratio of the $^{13}$CO and C$^{18}$O emission line peak main beam brightness on every pixel of the maps:
\begin{equation}
\frac{T\mathrm{_{MB}}(^{13}\mathrm{CO})}{T_\mathrm{MB}(\mathrm{C^{18}O})}=\frac{T\mathrm{_{ex,^{13}CO}}(1-e^{-\tau_{13}})}{T\mathrm{_{ex,C^{18}O}}(1-e^{-\tau_{18}})},
\label{optdepth}
\end{equation}
where $T_{\rm ex}$ is the excitation temperature of the isotopologues. The ratio of the optical depths is given by the abundance ratio of the two isotopologues: $\tau_{13}$/$\tau_{18}$\,=\,8 \citep{wilson1999}. The excitation temperatures are assumed to be approximately equal.

From the radiative transfer equation we obtain
\begin{equation}
I_{\rm obs}=(e^{-\tau_{13}}-1)I_{\rm bg}+(1-e^{-\tau_{13}})B(T_{\rm ex}),
\label{temp}
\end{equation}
where $I_{\rm obs}$ is the observed intensity of the emission line and $I_{\rm bg}$ is the background intensity (corresponding to $T_{\rm bg}$\,=\,2.7\,K). The contribution of $I_{\rm bg}$ is negligible compared to the line intensity, thus under local thermodynamic equilibrium (LTE) conditions and when the $^{13}$CO emission is optically thick, the observed line intensity approaches the intensity emitted by a black-body with a temperature of $T_{\rm ex}$.

For deriving the column density the typically optically thin C$^{18}$O line is the best choice. We use the method described by \citet{scoville1986} to derive the column density of C$^{18}$O. If the populations of all energy levels of the molecule can be characterized by a single temperature $T_{\rm ex}$ derived from Eq. \ref{temp} and we assume that the vibrationally excited states are not populated and all energy levels are populated under LTE conditions, the total column density of the molecule is derived as
\begin{equation}
%N_J/N_{tot}=\frac{g_u}{Z}\exp\left[-\frac{W}{k_BT_{\rm ex}} \right]
N_{\rm C^{18}O}=\frac{3k_{\rm B}}{8\pi^3B\mu^2}\frac{e^{hBJ(J+1)/k_{\rm B}T_{\rm ex}}}{(J+1)}\frac{T_{\rm ex}+hB/3k_{\rm B}}{1-e^{-h\nu/k_{\rm B}T_{\rm ex}}}\int\tau_{18}(v)dv,
\end{equation}
where $B$ and $\mu$ are the rotational constant and permanent dipole moment of the molecule, $k_{\rm B}$ is the Boltzmann-constant, $h$ is the Planck constant and $J$ is the rotational quantum number of the lower state in the observed transition. For the J=1$-$0 transition of C$^{18}$O this gives
\begin{equation}
N_{\rm C^{18}O}=2.42\times10^{14}\int\frac{T_{\rm ex}+0.88}{1-e^{-5.27/T_{\rm ex}}}\tau_{18}(v)dv
\end{equation}

In the case of optically thin medium the total mass of C$^{18}$O can be determined using
\begin{equation}
%F_{\rm ul}=\frac{N h\nu A_{\rm ul} X_{\rm u}}{4\pi d^2}
M_{\mathrm{C^{18}O}}=\frac{4\pi d m_\mathrm{C^{18}O}F_{\rm ul}}{h\nu_{\rm ul}A_{\rm ul}X_{\rm u}}
\end{equation}
where $d$ is the distance of the source, $m_{\rm C^{18}O}$ is the mass of the C$^{18}$O molecule, $F_{\rm ul}$ is the observed integrated flux density, $\nu_{\rm ul}$ is the rest frequency of the transition between the $u$ upper level and $l$ lower levels, A$_{\rm ul}$ is the Einstein-coefficient of the transition and $X_{\rm u}$ is the fractional population of the upper level (the ratio of the C$^{18}$O(1-0) molecules to the total number of C$^{18}$O molecules). In case of LTE, the level populations are thermalized and determined by the Boltzmann equation, thus $X_{u}$ can be computed if the temperature of the gas is known (T$_{\rm ex}$ from Eq. \ref{temp}). After deriving the C$^{18}$O mass, the $M_{H_2}$ hydrogen mass can be calculated using the canonical abundances of [$^{12}$CO]/[C$^{18}$O]\,=\,560 and [H$_2$]/[$^{12}$CO]\,=\,10$^4$. 

Due to the high signal-to-noise (S/N) of the observations the formal uncertainty of the calculated masses are small, but there are several systematic factors affecting the results. The error in the target distances is around 10$-$20\% and the used temperature values can have around a factor of two uncertainties, since we assume homogeneous temperature distribution. In the case of masses derived from the continuum (see Sect. \ref{sectcont}) the applied dust opacity coefficient may also have a factor of 2$-$4 uncertainty.

\subsection{Continuum}
\label{sectcont}

The 2.7\,mm continuum maps of the targets are shown in Fig. \ref{continuum}. The rms noise in the maps varies between 30$-$120 $\mu$Jybeam$^{-1}$. A single, roughly circular, compact source was detected at the optical position of V1057\,Cyg with 70$\sigma$. The continuum emission of V2492\,Cyg was detected with 79$\sigma$ with fainter emission extending to the southwest and another unresolved source with 16$\sigma$. We also detected V1735\,Cyg with 38$\sigma$ and fainter emission extending around it to the north. Another 9$\sigma$ point source appears at 12$\arcsec$ to the north. Four sources around V2493\,Cyg and three bright sources around V710\,Cas were observed but emission centered on the stars themselves was not detected. Only weak, 3$\sigma$ emission was detected close to the optical positions of V1515\,Cyg and V733\,Cep. 

We derived the parameters of the sources by fitting 2D Gaussian functions using the software package CASA \citep{mcmullin2007}. The resulting parameters are listed in Table \ref{continuum_data} where the 3$\sigma$ upper limits of the peak fluxes of the undetected sources are given as well. V1057\,Cyg and V2492 Cyg are resolved, while V1735\,Cyg is only marginally resolved in one direction with $b_{\rm deconv}$\,=\,0.9\,$\pm$\,0.4\,$\arcsec$. The rest of the continuum sources are resolved, except MMS2, located close to V2493 Cyg. The detected FUors have deconvolved sizes between 500$-$1800\,AU. The measured size and integrated flux of V2492\,Cyg is close to the one derived by \citet{hillenbrand2013} who gave a deconvolved size of 2.7$\arcsec\times$\,1.5$\arcsec$ with a position angle of 66$\degree$ and an integrated flux of 5.6\,$\pm$\,0.26\,Jy for their source. \citet{kospal2016} give a detailed description of the observed continuum sources around V2493\,Cyg. Both our line and continuum observations for V2493\,Cyg were partly published there and will only be discussed shortly here.

The mass of the continuum sources was derived using the equation
\begin{equation}
M_{\rm cont}=\frac{gS_{\nu}d^2}{\kappa_{\nu}B_{\nu}(T)},
\end{equation}
where $g$\,=\,100 is the gas-to-dust ratio, $S_{\nu}$ is the measured flux density at 2.7\,mm, $d$ is the distance, $\kappa_{\nu}$=0.2\,cm$^2$g$^{-1}$ is the dust opacity coefficient at 2.7\,mm based on \citet{ossenkopf1994} and $B_{\nu}$($T$) is the Planck function for a black-body with a temperature of $T$. The derived masses of the continuum sources coinciding with the targeted FUors using T\,=\,30\,K are in Table \ref{continuum_data}.

\subsection{$^{13}$CO(1-0) and C$^{18}$O(1-0)}
\label{sectline}

Both $^{13}$CO and C$^{18}$O emission were detected at each target with high S/N. The averaged spectra integrated over one synthetic beam at the position of the FUors (the center of the maps) are shown in Fig. \ref{avespec} and the parameters of these spectra are listed in Table \ref{avespec_data}. The channel maps are shown in the Appendix. The rms noise per channel ranges from $\sigma$\,=\,0.008$-$0.04\,Jybeam$^{-1}$. 

The $^{13}$CO line has a symmetric, Gaussian-like shape towards V1057\,Cyg and V2492\,Cyg, while it seems to be somewhat self-absorbed towards V733\,Cep. The spectra taken towards the other sources show multiple peaks that are either caused by self-absorption or might correspond to different line components in the line of sight, blended together. These blended line components also appear in the C$^{18}$O spectra. We find the broadest $^{13}$CO lines at the position of V710\,Cas with $\Delta v$\,$\approx$\,3.8\,kms$^{-1}$ and the smallest $^{13}$CO linewidth at V2492\,Cyg with $\Delta v$\,$\approx$\,1.2\,kms$^{-1}$. Apart from the bright, main $^{13}$CO line component we detected another velocity components in the direction of V1515\,Cyg and V2492\,Cyg, at 11.9\,kms$^{-1}$ and at 2.1\,kms$^{-1}$, respectively. These line components are not well visible in Fig. \ref{avespec}, because they mainly originate from a different region, not the central beam.

\begin{figure*}
\includegraphics[width=\linewidth, trim=0 0 0 0]{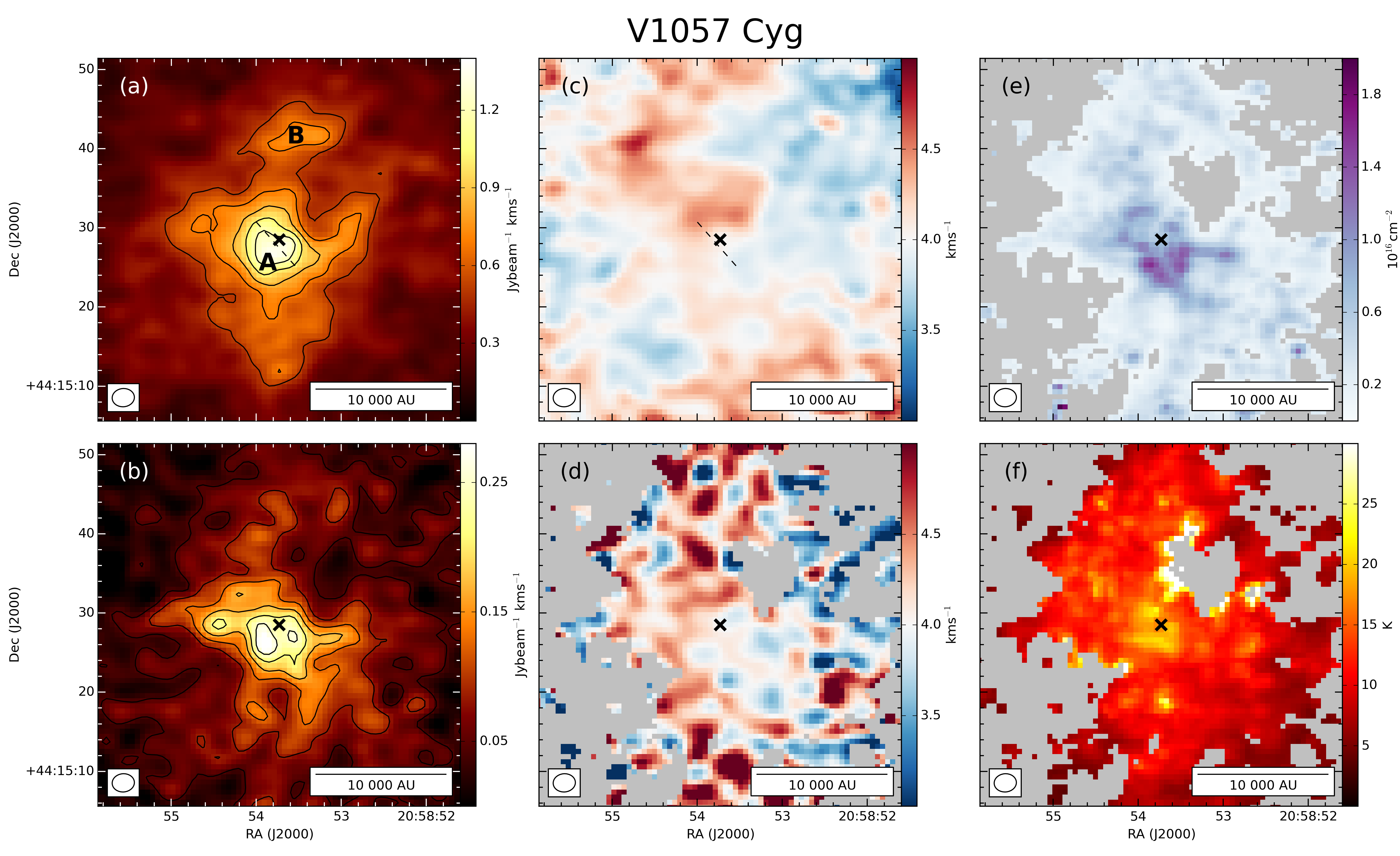}
	\caption{Moment and physical parameter maps of V1057\,Cyg: (a) $^{13}$CO integrated intensity; (b) C$^{18}$O integrated intensity; (c) $^{13}$CO intensity weighted velocity; (d) C$^{18}$O intensity weighted velocity; (e) C$^{18}$O column density; (f) temperature. Black cross marks the position of the FUor and letters indicate the clumps that are discussed in the text. The solid contours mark the $n\sigma_{\rm int}$ levels (the multiples of the noise level on the map); $\sigma_{\rm int}$\,=\,0.17\,Jybeam$^{-1}$\,kms$^{-1}$ and $n$\,=\,3, 4, ..., 8 on (a) and $\sigma_{\rm int}$\,=\,0.014\,Jybeam$^{-1}$\,kms$^{-1}$ and $n$\,=\,3, 6, ..., 18 on (b). The pixels with C$^{18}$O peak values less than 9$\sigma$ are coloured grey on panel (d), (e) and (f).}
	\label{v1057_fig}
\end{figure*}

\begin{figure}
\includegraphics[width=\linewidth, trim=0 0 0 0]{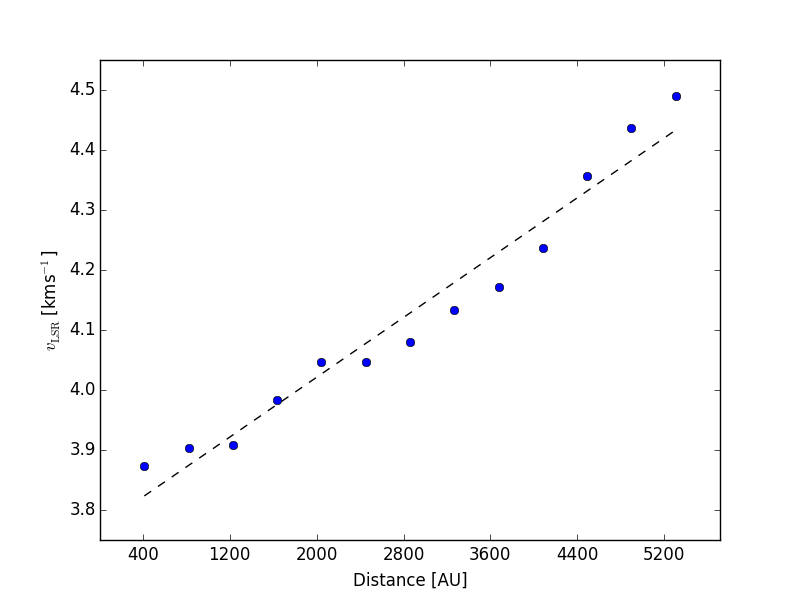}
	\caption{$^{13}$CO(1-0) $v_{\rm LSR}$ gradient across clump A of V1057\,Cyg along the line marked on panel (a) and (c) in Fig. \ref{v1057_fig}. The dashed line shows a linear fit to the values.}
	\label{v1057_grad}
\end{figure}

The integrated intensity (zeroth order moment) and the intensity weighted average velocity (first order moment) maps towards each target are plotted in panels (a)$-$(d) in Figs. \ref{v1057_fig}$-$\ref{v710_fig}. The $^{13}$CO and C$^{18}$O integrated intensity peaks are generally close to the optical stellar positions but considering the pointing accuracy of $\approx$\,0.5$\arcsec$ of the observations, only the optical position of V1735\,Cyg coincides with the detected $^{13}$CO peak and the optical position of V2492\,Cyg coincides with the C$^{18}$O peak. The pointing accuracy was calculated from the beam size, the S/N and the absolute positional accuracy of the telescope. V1057\,Cyg and V1735\,Cyg show a morphology with one strong, distinct integrated intensity peak at the center, but many smaller local peaks or clumps are found in the region around other targets, e.g. V1515\,Cyg or V2493\,Cyg, while V733\,Cep and V710\,Cas do not seem to coincide with CO emission peaks.

Using Eq. \ref{optdepth} to calculate optical depth, the $^{13}$CO emission was found mostly optically thick around the targets with values up to 15, but smaller areas with $\tau_{13}$ lower than unity can also be observed. The C$^{18}$O emission is generally optically thin with regions around V1515\,Cyg and V733\,Cep having $\tau_{18}$ somewhat higher than unity. The temperature distribution around the objects, as calculated from Eq. \ref{temp}, generally peaks close to the position of the FUors e.g. at V1057\,Cyg or V1515\,Cyg, but there are sources with complicated temperature distributions, e.g. V2493\,Cyg or V710\,Cas. The peak temperatures are between 15$-$75\,K with the less dense material showing 5$-$25\,K. The C$^{18}$O column densities are above 10$^{14}$\,cm$^{-2}$ around all our targets, corresponding to H$_2$ column densities of higher than 5.6\,$\times$\,10$^{20}$\,cm$^{-2}$, using the previously cited abundancies. The highest C$^{18}$O column density peaks appear at V710\,Cas with 6$-$7\,$\times$\,10$^{16}$\,cm$^{-2}$ and the lowest is at V1515\,Cyg and V733\,Cep with 7\,$\times$\,10$^{15}$\,cm$^{-2}$. The C$^{18}$O column density and the temperature maps are plotted in panels (e) and (f) in Figs. \ref{v1057_fig}$-$\ref{v710_fig}. In the regions where the C$^{18}$O opacities are approaching zero or there is higher noise in the C$^{18}$O spectra, undefined optical depths, temperatures and densities appear; these regions are masked out and plotted with grey in the figures.
 
Previous interferometric observations of V1057\,Cyg, V1515\,Cyg and V1735\,Cyg were presented by \citet{kospal2011}. Those $^{13}$CO(1-0) observations were made with the PdBI in 1993, using 4 antennas in the 4D1 configuration with baselines ranging from 24 to 64\,m. The synthesized beam was approximately 7$\arcsec\times$\,6$\arcsec$ and the rms noise per channel was around 0.15\,Jybeam$^{-1}$. The 2.7\,mm continuum was also measured towards the targets. V1057\,Cyg and V1515\,Cyg were not detected in those continuum measurements, while here we did detect both. They observed somewhat different morphologies around the targets but the discrepancies originate from the lower spatial resolution, the higher noise level and the sparser $uv$-coverage of the 1993 observations and will be discussed later as well. 

\section{Analysis of individual objects} 
\label{sectanalysis}

\subsection{V1057\,Cyg}

\begin{figure*}
\includegraphics[width=\linewidth, trim=0 0 0 0]{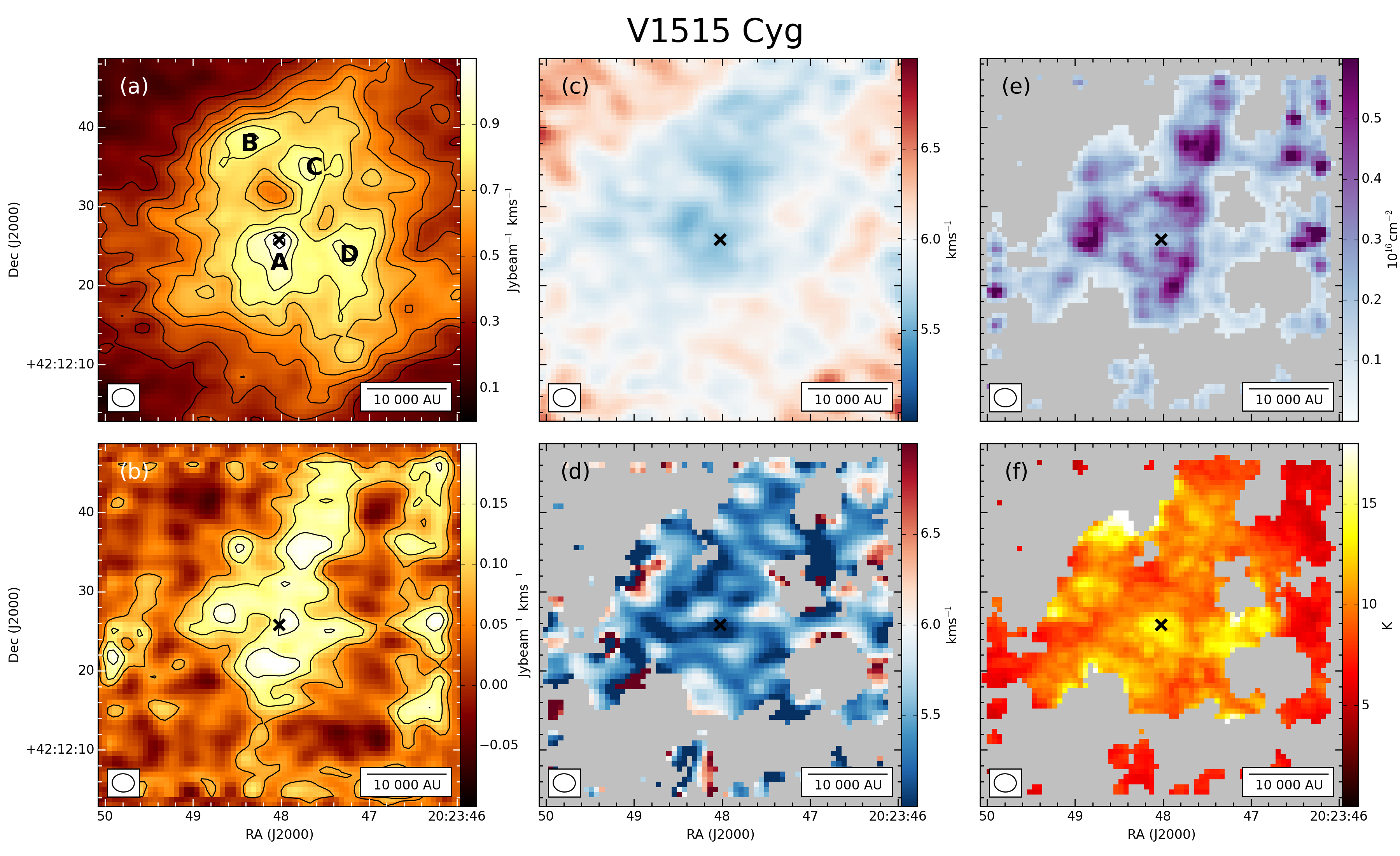}
	\caption{Moment and physical parameter maps of V1515\,Cyg. The panels are the same as in Fig. \ref{v1057_fig}, the position of the star and the clumps marked the same. $\sigma_{\rm int}$\,=\,0.1\,Jybeam$^{-1}$\,kms$^{-1}$ and $n$\,=\,3, 4, ..., 12 on (a) and $\sigma_{\rm int}$\,=\,0.024\,Jybeam$^{-1}$\,kms$^{-1}$ and $n$\,=\,3, 5, 7 on (b). The pixels with C$^{18}$O peak values less than 9$\sigma$ are coloured grey on panel (d), (e) and (f).}
	\label{v1515_fig}
\end{figure*}

The $^{13}$CO and C$^{18}$O lines towards the FUor are single-peaked, without significant self-absorption. Some excess emission on the $^{13}$CO blue line wing can be detected between 0$-$3.4\,kms$^{-1}$. The integrated intensity map of $^{13}$CO shows one strong peak at 1.5$\arcsec$ to the south from the center with 1.3\,Jybeam$^{-1}$\,kms$^{-1}$ (see Fig. \ref{v1057_fig}a). The central clump around this peak (clump A) is roughly circular, with a radius of 5$\arcsec$, and it appears in C$^{18}$O as well. It also roughly corresponds to the detected continuum source but it is centered more to the southeast. Two protrusions to the east and to the west from clump A can be observed in the integrated intensity of both CO isotopologues. The western protrusion extends towards the nebulosity "arm" seen on optical images \citep{duncan1981}. Integrating the excess $^{13}$CO emission between 0$-$3.4\,kms$^{-1}$ also shows this western protrusion and additionally, clump B. A $^{13}$CO and C$^{18}$O $v_{\rm LSR}$ gradient can be observed in clump A in the southwest-northeast direction (Fig. \ref{v1057_fig}c), which can be seen on the channel maps as well %(Figs. \ref{v1057_chan1} and \ref{v1057_chan2}) 
where clump A appears at around 3.33\,kms$^{-1}$ to the south from the center then moves towards the east between 4.17 and 4.49\,kms$^{-1}$. The $^{13}$CO velocities along the line marked in panels (a) and (c) in Fig. \ref{v1057_fig} are plotted and fitted with a line in Fig. \ref{v1057_grad}. The velocity gradient is 0.08\,kms$^{-1}$\,arcsec$^{-1}$ or 27.6\,kms$^{-1}$\,pc$^{-1}$ at a distance of 600\,pc. It is also apparent that at lower velocities the more diffuse emission in the channel maps is elongated to the west-east direction but it becomes elongated in the north-south direction at higher velocities, where clump B also merges with the central region. The C$^{18}$O channel maps %(Fig. \ref{v1057_chan2}) 
are very similar to the $^{13}$CO at corresponding velocities, which suggests that the emission of the two molecules originate in the same volume of material. 

The calculated temperatures are higher in clump A (20$-$22\,K) than in the more diffuse region outside of it. This suggests heating from the central source, although there is a local minimum at the position of the FUor with 17\,K inside clump A. The $^{13}$CO emission is only marginally optically thick inside clump A with $\tau_{13}$\,=\,1$-$3. The C$^{18}$O column density is above 10$^{16}$\,cm$^{-2}$ in the southern part of clump A with a peak of 1.6\,$\times$\,10$^{16}$\,cm$^{-2}$ and there are at least three arcs of denser material around this region. In the northern half of clump A however, the density decreases to 4.5\,$\times$\,10$^{15}$\,cm$^{-2}$. The mass of clump A using an average $T$\,=\,16.4\,K is 0.21\,$M_{\odot}$, which is almost twice the value derived by \citet{kospal2011} from $^{13}$CO emission. Assuming that the $^{13}$CO emission is optically thick and deriving the mass from it we get 0.14\,$M_{\odot}$, which is more consistent with their result, implying that the difference originates from $^{13}$CO being optically thick. The temperature peak measured close to the FUor and the detected velocity gradient might signal a rotating envelope with a radius of 5$\arcsec$ (3000\,AU).
 
\subsection{V1515\,Cyg}

\begin{figure*}
\includegraphics[width=\linewidth, trim=0 0 0 0]{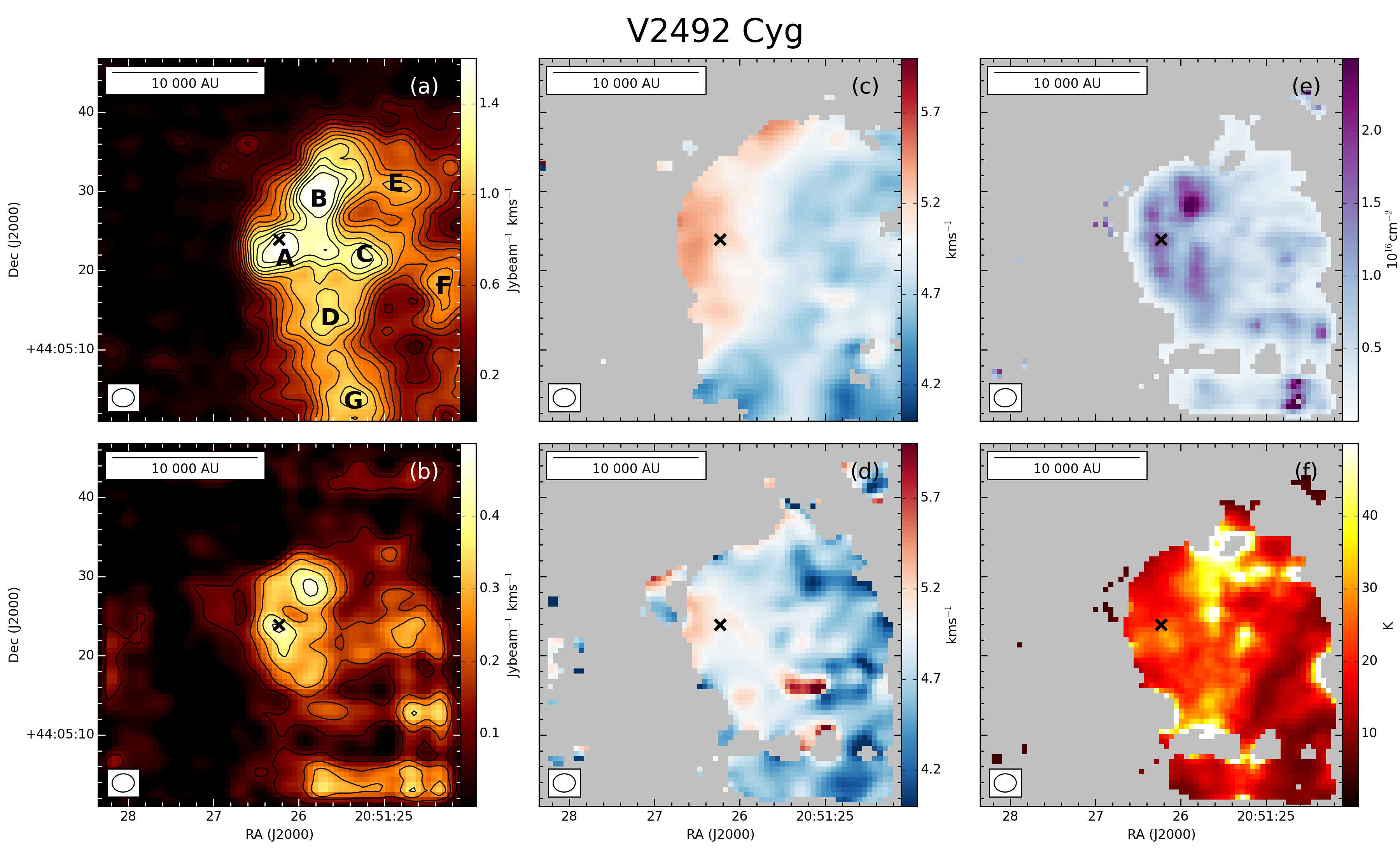}
	\caption{Moment and physical parameter maps of V2492\,Cyg. The panels are the same as in Fig. \ref{v1057_fig}, the position of the star and the clumps marked the same. $\sigma_{\rm int}$\,=\,0.04\,Jybeam$^{-1}$\,kms$^{-1}$ and $n$\,=\,3, 6, ..., 36 on (a) and $\sigma_{\rm int}$\,=\,0.03\,Jybeam$^{-1}$\,kms$^{-1}$ and $n$\,=\,3, 6, ..., 18 on (b). The pixels with C$^{18}$O peak values less than 9$\sigma$ are coloured grey on panel (d), (e) and (f).}
	\label{v2492_fig}
\end{figure*}

The $^{13}$CO line of V1515\,Cyg shows two peaks with possible self absorption in the mapped area with no significant excess on the line wings. The brightest peak of the C$^{18}$O line is at the same velocity as the $^{13}$CO peak and traces of another line component also appear. Additionally, apart from this bright line at 5.7\,kms$^{-1}$, there is another, faint $^{13}$CO velocity component at around 11.8\,kms$^{-1}$. Its integrated intensity shows a flaring structure that covers the northwest region of the map. The integrated intensity map of the brighter line shows a clumpy, extended, roughly spherical structure (Fig. \ref{v1515_fig}a) that fills the primary beam. The maximum is inside clump A, with 1.1\,Jybeam$^{-1}$\,kms$^{-1}$, roughly coinciding with the FUor. Three other clumps (B, C and D) are seen at 9 and 13.5$\arcsec$ to the northwest and at 8.7$\arcsec$ to the west. A hole in the emission appears between clump A and C. The overall distribution of stronger and weaker peaks actually correspond well to the arc-like shape detected by \citet{kospal2011} considering the lower resolution of their measurement. This arc-like shape is oriented similarly as the arc of nebulosity on optical images. No clear velocity gradient can be defined.

Looking at the $^{13}$CO channel maps %(Fig. \ref{v1515_chan1}) 
first we see clump A appearing at low velocities, then gradually many smaller clumps emerge around it in a ring. These clumps fragment, merge and change positions slightly throughout the measured velocity interval. Clump A disappears around 5.63\,kms$^{-1}$ but some of the smaller clumps are visible at even 5.9\,kms$^{-1}$. Above 6\,kms$^{-1}$ only diffuse, extended emission with small local peaks is present. This morphology may suggest an expanding bubble around the star, where at low velocities we detect the approaching wall of the structure, at medium velocities the ring-like circumference, then at even higher velocities the far side of the bubble wall moving away from the observer. If the structure is indeed expanding, the velocity differences and the size of the ring would suggest a dynamical age of $\approx$\,32\,000 years.

The C$^{18}$O integrated intensity map is somewhat different from the $^{13}$CO emission, showing an elongated structure in the northwest-southeast direction. Three strong peaks (around 0.2\,Jybeam$^{-1}$\,kms$^{-1}$) can be seen at 7$\arcsec$ to the southeast, at 11$\arcsec$ to the northwest, and close to the center. Similarly in the C$^{18}$O channel maps%(Fig. \ref{v1515_chan2})
, there are only 3$-$4 larger clumps present between 5.1 and 5.32\,kms$^{-1}$, two of these approximately coincide with clumps A and C. There are another 3$-$4 clumps at the western edge of the map between 5.53 and 5.85\,kms$^{-1}$, where the $^{13}$CO emission is much less significant at these velocities. 

\begin{figure*}
\includegraphics[width=\linewidth, trim=0 0 0 0]{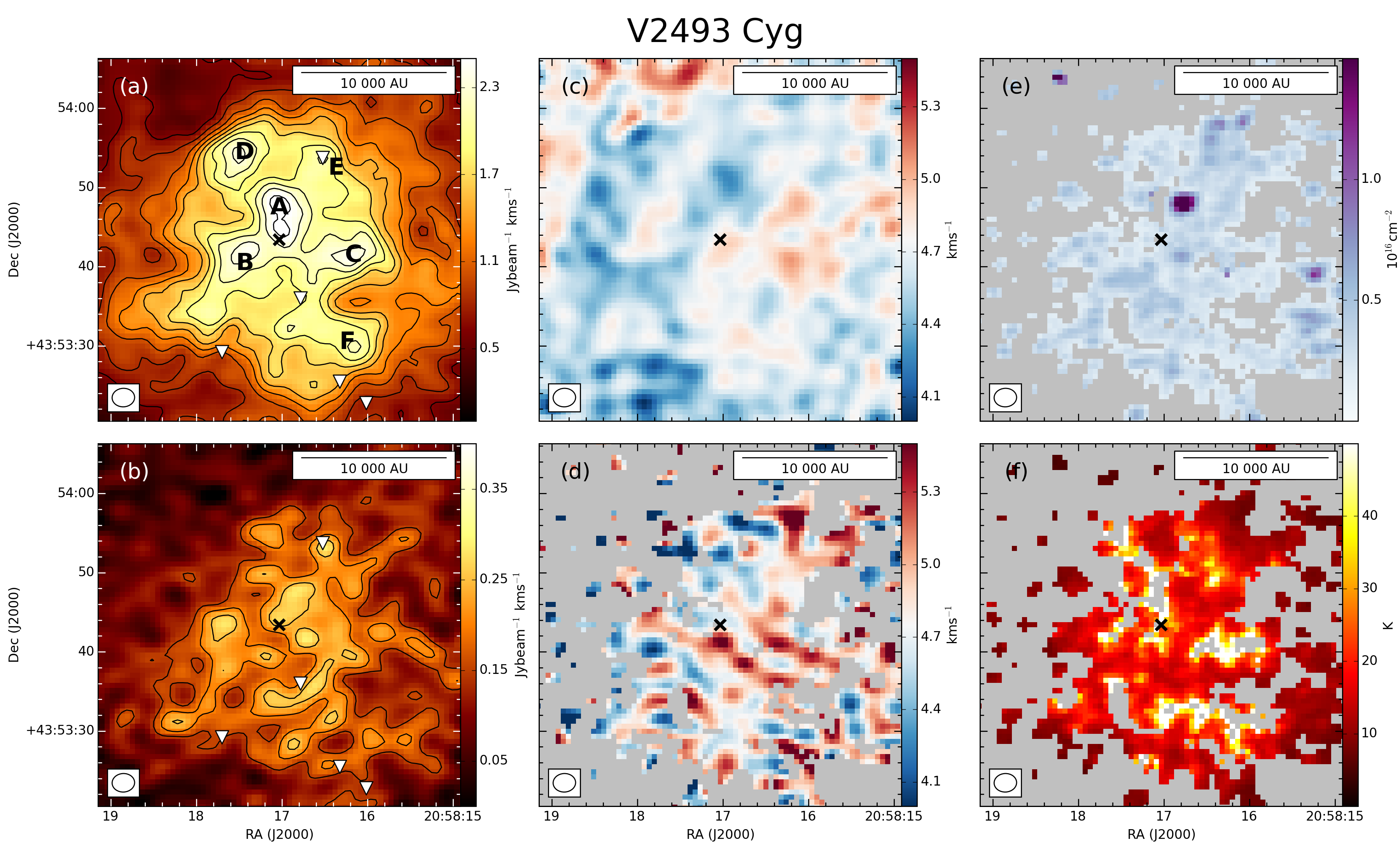}
	\caption{Moment and physical parameter maps of V2493\,Cyg. The panels are the same as in Fig. \ref{v1057_fig}, the position of the star and the clumps marked the same. $\sigma_{\rm int}$\,=\,0.21\,Jybeam$^{-1}$\,kms$^{-1}$ and $n$\,=\,3, 4, ..., 12 on (a) and $\sigma_{\rm int}$\,=\,0.05\,Jybeam$^{-1}$\,kms$^{-1}$ and $n$\,=\,3, 4, ..., 8 on (b). The triangles mark the millimeter sources from \citet{dunham2012} as in Fig. \ref{continuum}. The pixels with C$^{18}$O peak values less than 7$\sigma$ are coloured grey on panel (d), (e) and (f).}
	\label{v2493_fig}
\end{figure*}

The high density regions in Fig. \ref{v1515_fig}e coincide with the positions of the clumps that appear both in some of the $^{13}$CO and C$^{18}$O channel maps. In these clumps the C$^{18}$O emission becomes optically thick. The $N_{\rm C^{18}O}$ distribution peaks in clump C with 5\,$\times$\,10$^{15}$\,cm$^{-2}$ and in the other clumps with $\approx$\,4.5\,$\times$\,10$^{15}$\,cm$^{-2}$. The temperature distribution peaks at the center in clump A with 14.7\,K, while most of the mapped area has temperatures around 8$-$10\,K. The mass of clump A with a radius of 5$\arcsec$ and an average temperature of 11.2\,K is 0.26\,$M_{\odot}$ from C$^{18}$O flux density integrated between 4 and 5.53\,kms$^{-1}$, where clump A is present in the channel maps. This is lower than the value of 0.42\,$M_{\odot}$ derived by \citet{kospal2011} using $^{13}$CO, which again may originate from $^{13}$CO optical depth effects. We note however that the previous observations did not resolve the structure around V1515\,Cyg into clumps. The three larger clumps in the $\tau_{13}$ ring have average temperatures of 9$-$11\,K and masses of 0.04$-$0.05\,$M_{\odot}$.

\subsection{V2492\,Cyg}

The $^{13}$CO line in V2492\,Cyg is single-peaked and shows no significant self-absorption or line wing excesses. Apart from the brighter line around 5\,kms$^{-1}$, there is another faint $^{13}$CO velocity component at 2\,kms$^{-1}$. The integrated intensity of this component shows emission at the center and in a point source-like peak inside diffuse emission at 22$\arcsec$ to the northeast from the center. The integrated intensity map of the brighter $^{13}$CO line component shows an asymmetric, north-south oriented morphology with emission missing to the northeast, east and southeast (Fig. \ref{v2492_fig}a). The primary peak is at 1.3$\arcsec$ to the west from the center with 1.7 Jybeam$^{-1}$\,kms$^{-1}$ in clump A. The similarly bright clump B is found at 7.3$\arcsec$ to the northwest and the smaller clump C is inside a protrusion at 10$\arcsec$ to the west. Two other small clumps are located close to the western edge of the map (E and F). The more diffuse emission reaches beyond the map edge towards the south, containing at least two other clumps (D and G). The C$^{18}$O line integrated intensity distribution shows the same features but with only two significant clumps in the center region, that coincide with clumps A and B with peaks of 0.46 and 0.48\,Jybeam$^{-1}$\,kms$^{-1}$. The smaller clumps at the western and southern edges of the map also appear.

The overall distribution of the emission seems to be moving from the southwest to the northeast on the channel maps of both isotopologues, and this appears as a large-scale velocity gradient of 0.05\,kms$^{-1}$\,arcsec$^{-1}$ (18.7\,kms$^{-1}$\,pc$^{-1}$ at 550\,pc) across the position of the star from the west to the east (Fig. \ref{v2492_fig}c and d). At low velocities there is emission only to the southern-southwestern edge of the map in the $^{13}$CO channel maps%(Fig. \ref{v2492_chan1})
, but above 4.18\,kms$^{-1}$ four clumps form in a linear, northwest-southeast configuration. The two northern and two southern clumps merge, forming clumps A$-$D and at around these same channels clump E, F and G emerge as well. The clumps A-D stay in the same formation between 4.82 and 5.03\,kms$^{-1}$ and they start to merge into one structure above 5\,kms$^{-1}$, that is centered somewhat to the east from the center. Above 5.67\,kms$^{-1}$ only two bright clumps are present, clump A and B.  

The eastern edge of the elongated $^{13}$CO structure coincides with the bright rim of H$\alpha$ nebulosity observed by PTF \citep{hillenbrand2013}. The CO emission starts just inside the nebulosity and also coincides with the bright region around the FUor in $Herschel$ images \citep{kospal2013}. This structure may indicate a shock front edge, formed by the same ionizing radiation that causes the bright rim in H$\alpha$, compressing the material around V2492\,Cyg.

The densest region on the $N_{\rm C^{18}O}$ map is directly behind the shock front in an arc-like structure, including clump A and B with peaks of 1.6 and 2.6\,$\times$\,10$^{16}$\,cm$^{-1}$. Clump A also seems to be heated to 30\,K in contrast with the 20\,K temperatures on most of the mapped area. Two other warmer spots appear on the temperature maps with 55 and 70\,K, one roughly coincides with clump C and the other is in the middle of the triangle of clumps A, B and C. Clump A coincides with the strong continuum source that appears at 2.7\,mm. The "tail"-shaped continuum feature points towards clump C but does not extend that far. The other, faint continuum point source is located to the south from clump C. A third 2.7\,mm source was detected by \citet{hillenbrand2013} which does not appear in our continuum map but it would roughly coincide with our clump D on the CO maps. Clump B does not appear in the continuum, despite it being the hottest and densest source in our CO maps. Using 23.5\,K and 29.8\,K average clump temperatures and radii of 3.5$\arcsec$ we calculate the masses of clump A and B as 0.17 and 0.23\,M$_{\odot}$, while clump C has a mass of 0.09\,M$_{\odot}$, using 26.7\,K and 3$\arcsec$.

\subsection{V2493\,Cyg}

\begin{figure*}
\includegraphics[width=\linewidth, trim=0 0 0 0]{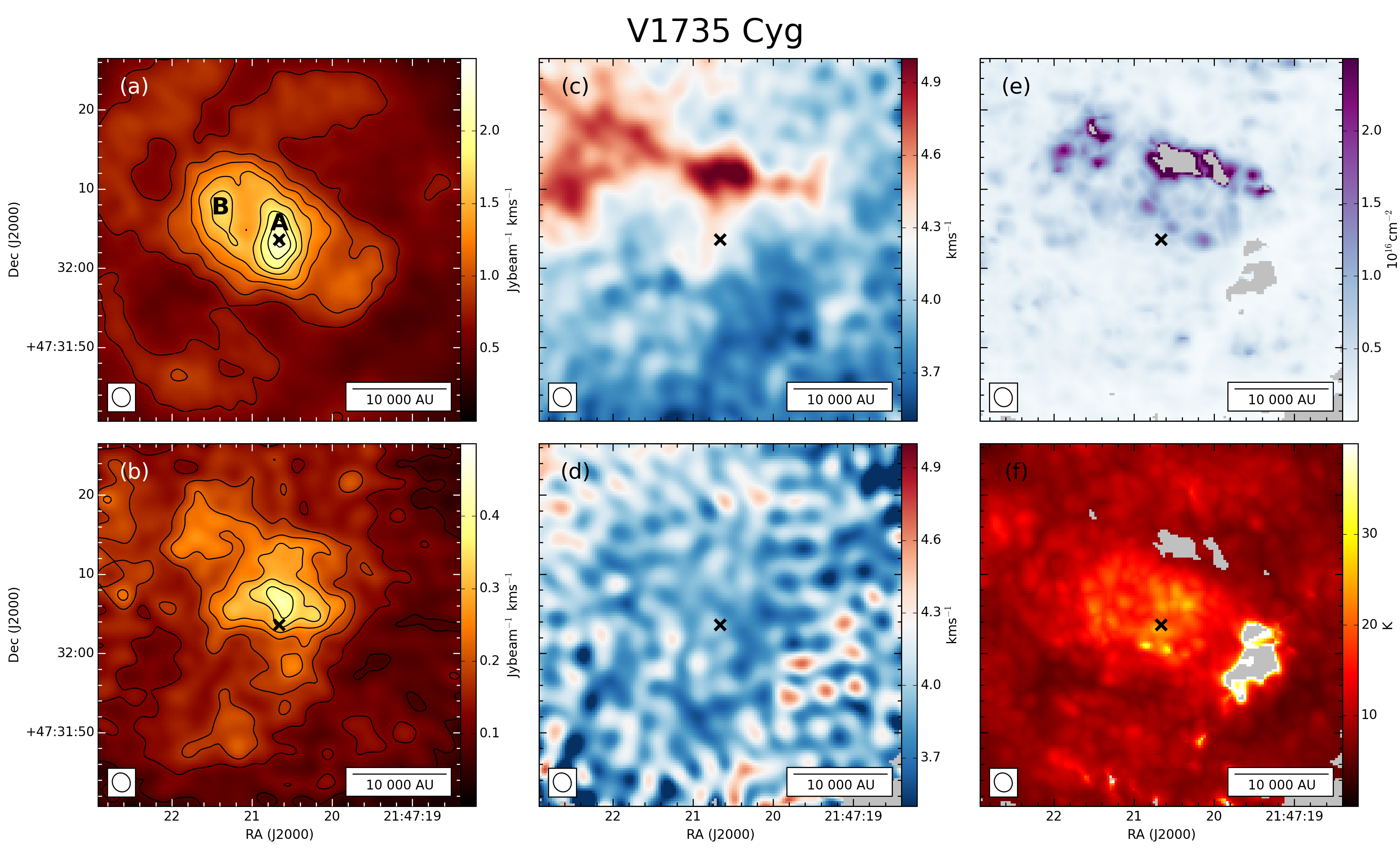}
	\caption{Moment and physical parameter maps of V1735\,Cyg. The panels are the same as in Fig. \ref{v1057_fig}, the position of the star and the clumps marked the same. $\sigma_{\rm int}$\,=\,0.23\,Jybeam$^{-1}$\,kms$^{-1}$ and $n$\,=\,3, 4, ..., 13 on (a) and $\sigma_{\rm int}$\,=\,0.025\,Jybeam$^{-1}$\,kms$^{-1}$ and $n$\,=\,3, 5, ..., 15 on (b). The pixels with C$^{18}$O peak values less than 18$\sigma$ are coloured grey on panel (d), (e) and (f).}
	\label{v1735_fig}
\end{figure*}

Both $^{13}$CO and C$^{18}$O show wide, multi-peaked lines and some excess mainly on the blue line wings. The $^{13}$CO integrated intensity map shows an extended, roughly circular, 20$\arcsec\times$\,30$\arcsec$ bright area around the star with several little clumps (Fig. \ref{v2493_fig}a). The elongated clump A has two peaks north from the center with 2.6\,Jybeam$^{-1}$\,kms$^{-1}$. There are three brighter (B$-$D) and two fainter clumps (E and F) around. These clumps appear above 2.6\,kms$^{-1}$ in the channel maps%(Fig. \ref{v2493_chan1})
, then fragment and merge throughout the whole velocity interval. Clump A is in fact two small clumps close to each other, the northern at lower velocities than the southern and this appears as a velocity gradient of 0.06\,kms$^{-1}$\,arcsec$^{-1}$ (22.5\,kms$^{-1}$\,pc$^{-1}$ at 550\,pc) in the clump. Clump E corresponds to the 2.7\,mm continuum source MMS1 \citep{dunham2012}. We also detected MMS2$-$MMS4 in our continuum maps but none of them correspond to any of the CO clumps. The FUor itself does not appear in the continuum and even in $^{13}$CO, emission centered on it can only be observed by integrating between 5.45 and 6.52\,kms$^{-1}$, as shown by \citet{kospal2016}. The C$^{18}$O integrated intensity map (Fig. \ref{v2493_fig}c) shows only weak emission, the peak is at 4$\arcsec$ to the southwest from the center with 0.3 Jybeam$^{-1}$\,kms$^{-1}$. The C$^{18}$O channel maps %(Fig. \ref{v2493_chan2})
are similar to the overall distribution of the $^{13}$CO emission. It is difficult to recognize clumps here, except between 4.82 and 5.03\,kms$^{-1}$ where a point-like, bright source appears at 6$\arcsec$ to the northwest from the center. This source is close to clump A but does not coincide with it.

As seen on Fig. \ref{avespec} the three peaks on the $^{13}$CO line profile are around the same velocities as the peaks on the C$^{18}$O line, thus they may be in fact three line components centered on 3.4, 5.1 and 7.1\,kms$^{-1}$. Integrating the emission between 0.3$-$4.3, 4.3$-$6.5 and 6.5$-$8.7\,kms$^{-1}$ it is apparent that the lower velocity line traces mainly clump A, B and D, the middle one clump A, C and E and the high velocity line component shows weaker, clumpy emission throughout the whole area. 

The derived temperatures are generally 10$-$20\,K, with higher peaks that is mainly due to the lower S/N of the C$^{18}$O spectra. The C$^{18}$O column densities are generally around 1$-$6\,$\times$\,10$^{15}$\,cm$^{-2}$ and rise to 2.7\,$\times$\,10$^{16}$\,cm$^{-2}$  at the point-source-like object to the northwest. Here even $\tau_{18}$ is close to unity. The mass of the clump around the FUor integrated between 5.45 and 6.52\,kms$^{-1}$ with a radius of 3$\arcsec$ is 0.024\,M$_{\odot}$. Since temperatures are not well measured around the center, we use here an average T\,=\,21.6\,K calculated inside the primary beam. The clumps A$-$D have masses of 0.05$-$0.15\,M$_{\odot}$ using the same temperature.

\begin{figure*}
\includegraphics[width=\linewidth, trim=0 0 0 0]{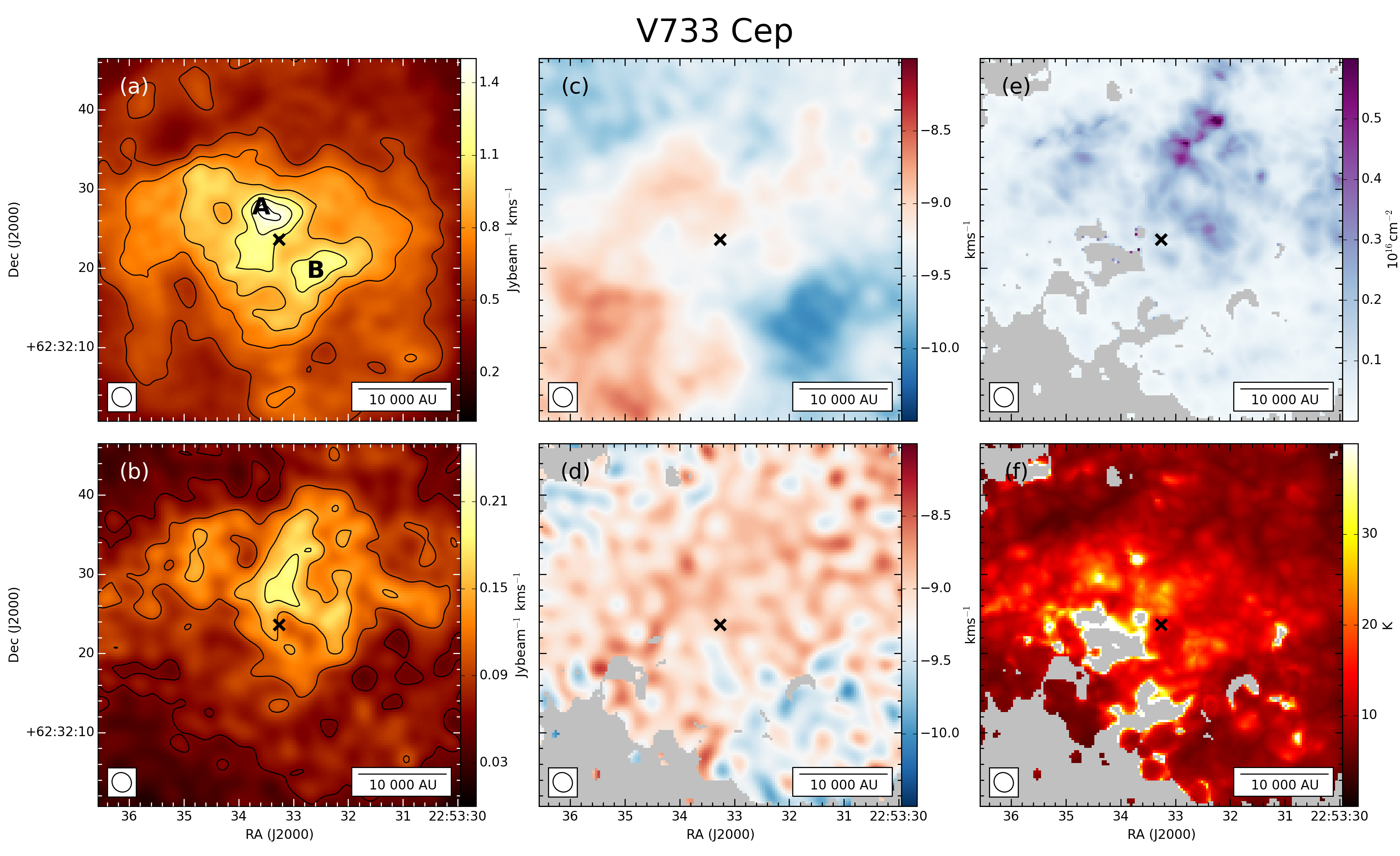}
	\caption{Moment and physical parameter maps of V733\,Cep. The panels are the same as in Fig. \ref{v1057_fig}, the position of the star and the clumps marked the same. $\sigma_{\rm int}$\,=\,0.18\,Jybeam$^{-1}$\,kms$^{-1}$ and $n$\,=\,3, 4, ..., 8 on (a) and $\sigma_{\rm int}$\,=\,0.02\,Jybeam$^{-1}$\,kms$^{-1}$ and $n$\,=\,3, 5, ..., 9 on (b). The pixels with C$^{18}$O peak values less than 10$\sigma$ are coloured grey on panel (d), (e) and (f).}
	\label{v733_fig}
\end{figure*}

\subsection{V1735\,Cyg}

Both the $^{13}$CO and C$^{18}$O lines in V1735\,Cyg show at least two peaks and excess emission on both line wings. In the $^{13}$CO integrated intensity map an elliptic, bright area appears that is elongated in the northeast-southwest direction (Fig. \ref{v1735_fig}a). The southeastern part of the structure is fainter than the northwestern. The emission peak (2.3\,Jybeam$^{-1}$\,kms$^{-1}$) coincides with the FUor inside a north-south elongated, bright, elliptic clump that we call clump A. There is a weaker clump to the northeast that protrudes from this central area (clump B). The point source detected in the 2.7\,mm continuum roughly coincides with clump A, but the weaker continuum features around the strong detection have no counterparts in the CO maps. In the $^{13}$CO channel maps %(Fig. \ref{v1735_chan1})
an S-shaped region of emission appears at the center between 2.41 and 3.15\,kms$^{-1}$ that breaks up into several clumps. Between 3.68 and 4.32\,kms$^{-1}$ there are three bright regions at the center, these form another bright S-shaped structure which then breaks up again: the southwestern and northeastern parts disappear by 4.64\,kms$^{-1}$ and only the central clump is significant at higher velocities. Weaker emission is present to the north even above 8\,kms$^{-1}$. The C$^{18}$O integrated intensity map (Fig. \ref{v1735_fig}c) is very different from $^{13}$CO, showing significant emission only on the north side of the star with a peak of 0.4\,Jybeam$^{-1}$\,kms$^{-1}$. The reason for this can be seen on the channel maps%(Fig. \ref{v1735_chan2})
, where the same S-shaped emission appears but the north-northwestern side of the structure is much brighter. Above 3.79\,kms$^{-1}$ only an irregularly shaped central source is present, somewhat to the northwest from the center. 

\begin{figure*}
\includegraphics[width=\linewidth, trim=0 0 0 0]{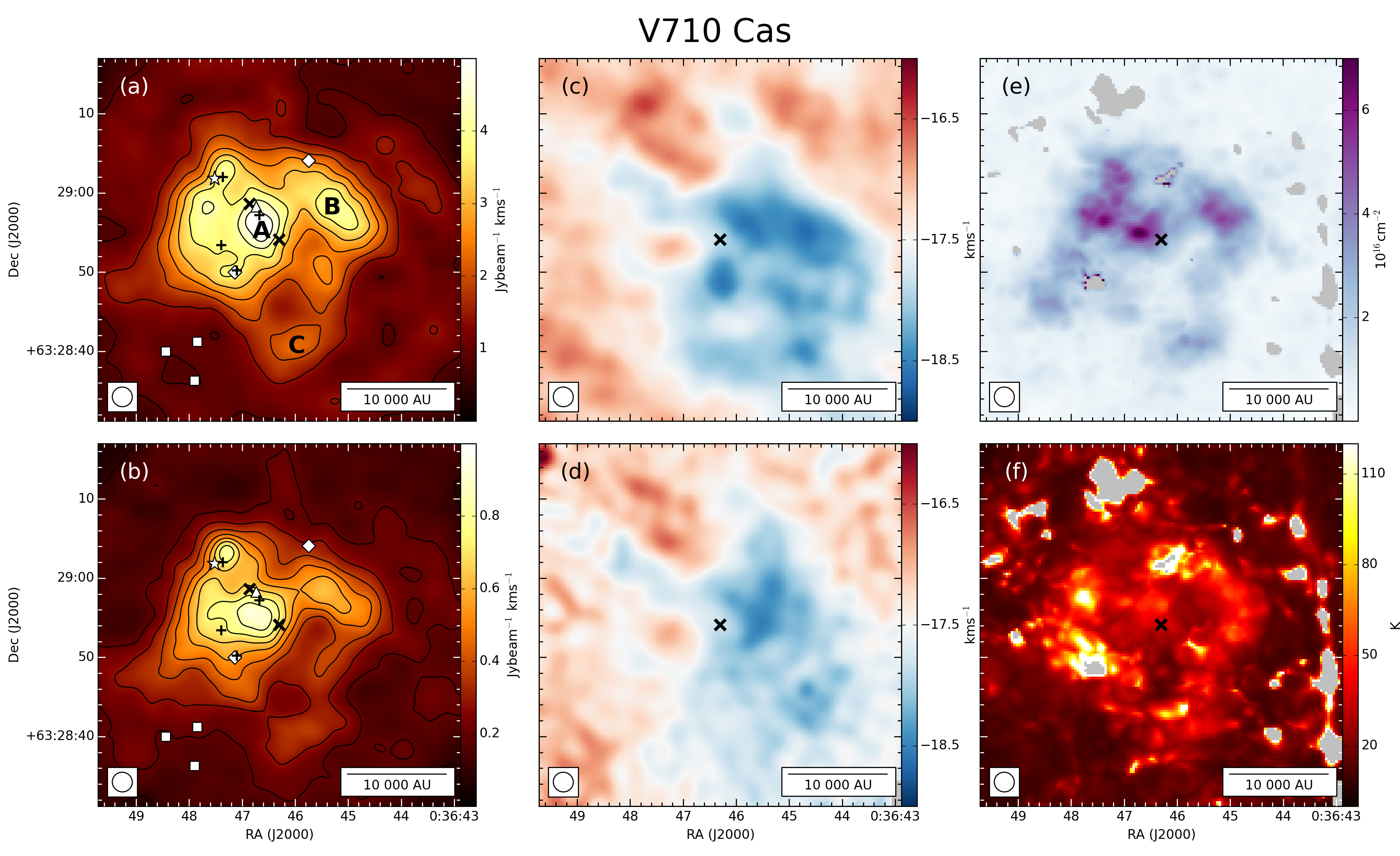}
	\caption{Moment and physical parameter maps of V710\,Cas. The panels are the same as in Fig. \ref{v1057_fig}, the position of the star and the clumps marked the same. $\sigma_{\rm int}$\,=\,0.28\,Jybeam$^{-1}$\,kms$^{-1}$ and $n$\,=\,3, 5, ..., 17 on (a) and $\sigma_{\rm int}$\,=\,0.031\,Jybeam$^{-1}$\,kms$^{-1}$ and $n$\,=\,3, 6, ..., 33 on (b). The previously detected sources are marked similarly as in Fig. \ref{continuum}. The pixels with C$^{18}$O peak values less than 18$\sigma$ are coloured grey on panel (d), (e) and (f).}
	\label{v710_fig}
\end{figure*}

The velocity map of $^{13}$CO (Fig. \ref{v1735_fig}c) shows somewhat higher values in clump A than in the rest of the elliptical structure and there is a high-velocity lobe to the north. This is caused by emission appearing at 8$-$9\,kms$^{-1}$ velocities in the channel maps. The linewidth distribution also shows two areas with broad lines (4.5$-$5.5\,kms$^{-1}$) to the northwest and to the southeast of the star. The northern one coincides with the northern end of the high-velocity lobe. Integrating the $^{13}$CO line wings between $-$0.2$-$2\,kms$^{-1}$ and 7$-$9.4\,kms$^{-1}$, it is apparent that the blue and red-shifted areas occupy two distinct regions to the south and north of the center, respectively (Fig. \ref{fuor_outflows}). The high linewidth regions are spatially connected to the lobes, suggesting the presence of an outflow. The orientation of the lobes agrees with the outflow morphology detected by \citet{evans1994} in $^{12}$CO, although their angular resolution was much lower and they mapped a larger area. The outflow was proposed to be driven by the other continuum source at 20$\arcsec$ to the northeast but the geometry of the lobes detected here suggests that it may be driven by V1735\,Cyg.

Around the region with high $^{13}$CO linewidths, the C$^{18}$O emission is very bright, while $^{13}$CO is very strongly self-absorbed, causing their line ratio to be higher than in the optically thick limit. This results in very high or undefinable optical depths in the area. The C$^{18}$O column densities are significantly higher here than to the south: it is generally around 0.5$-$2.3\,$\times$\,10$^{16}$\,cm$^{-2}$ with no clear peak at the center. The $^{13}$CO optical depth also shows somewhat higher values to the north (4$-$10) than to the south (1$-$4). The temperatures are around 15$-$20\,K in the elliptical structure and peak to the south from the center with 30 and 33\,K. We measure 25\,K at the FUor position. We calculated the dust optical depth from the continuum map by subtracting a Planck-function defined by the derived gas temperatures pixel-by-pixel and this shows that the strong continuum source is mainly a temperature effect and there is not a high-density structure in the map, similarly to the C$^{18}$O column density distribution. The gas-based mass of clump A with a radius of 4$\arcsec$ and an average temperature of 19.7\,K is 0.51\,M$_{\odot}$.

\subsection{V733\,Cep}

Our observations of V733\,Cep show self-absorbed $^{13}$CO emission with some excess on the blue line wing. The $^{13}$CO integrated intensity peak is located at 4$\arcsec$ to the northeast from the FUor, with 1.5\,Jybeam$^{-1}$\,kms$^{-1}$, in clump A (Fig. \ref{v733_fig}a). There is another, fainter clump to the southwest, clump B. In the $^{13}$CO channel maps at low velocities there is only significant emission in the southwestern corner, then an elongated structure appears in the northeast-southwest direction, between $-$10.2 and $-$9.77\,kms$^{-1}$. Clump A appears around $-$9.77\,kms$^{-1}$ and it is significant until $-$8.08\,kms$^{-1}$, somewhat changing its position and size. Clump B is present in these channels as well. Other small clumps appear further northeast, to the south and to the southwest. At $-$8.71\,kms$^{-1}$ there are only three strong clumps and above $-$7.86\,kms$^{-1}$ there is no significant emission in the center. However, even at $-$6\,kms$^{-1}$ there is emission in the southeastern region of the map. There is no clear systematic change in the line velocities but $^{13}$CO bright areas generally have higher velocities and smaller line widths (2$-$2.5\,kms$^{-1}$) than the background. 

The C$^{18}$O line integrated intensity map shows emission mainly to the north of the source, since between $-$9.35\,kms$^{-1}$ and $-$8.5 the strongest C$^{18}$O emission comes from the brightest $^{13}$CO clumps, to the northeast, north and northwest. Similarly to V1515\,Cyg the $^{13}$CO channel maps might be interpreted as a small expanding shell with the receding side seen around $-$9.77\,kms$^{-1}$ and the approaching wall around $-$8.08\,kms$^{-1}$. This is supported by the fact that there is no C$^{18}$O emission peak towards the FUor in any of the channels. 

The temperatures are generally higher inside the bright $^{13}$CO emission areas than at the edges of the map, with 10$-$20\,K and peaks of 45 and 75\,K to the northeast. The high opacity filaments to the north seem to be cold, around 10\,K. We measure higher $N_{\rm C^{18}O}$ values to the north than the south, with peaks of 4.8 and 7\,$\times$\,10$^{15}$\,cm$^{-2}$. Clump A with an average T\,=\,27.3\,K has a mass of 0.22\,M$_{\odot}$ and clump B with T\,=\,17.9\,K has a mass of 0.06\,M$_{\odot}$.

Integrating the $^{13}$CO emission in the line wings between $-$12.9 and $-$10.7, then $-$7.4 and $-$6\,kms$^{-1}$ (see Fig. \ref{fuor_outflows}) the blue wing shows emission mostly to the southwest and the red wing to the southeast, in a somewhat clumpy structure. If this indeed indicates an outflow, the source that drives it should be located around 20$\arcsec$ to the south of the center, rather than V733\,Cep itself.

\subsection{V710\,Cas}

Both the $^{13}$CO and C$^{18}$O lines towards V710\,Cas (RNO\,1B) show multiple peaks and possibly self-absorption. The $^{13}$CO integrated intensity map has a bright peak at 2.5$\arcsec$ to the northeast of RNO\,1B with 4.6\,Jybeam$^{-1}$\,kms$^{-1}$ (Fig. \ref{v710_fig}a). This circular bright area that we call clump A is located between RNO\,1B and C. Further peaks can be found at the end of an arc-like structure to the west (clump B), little local peaks to the east and the much fainter clump C to the south. The C$^{18}$O integrated intensities show the same morphology with a peak of 1\,Jybeam$^{-1}$kms$^{-1}$ at 3.8$\arcsec$ from RNO\,1B that roughly coincides with the $^{13}$CO peak. Both the $^{13}$CO and the C$^{18}$O velocity maps show clump A and the eastern side of the arc at higher velocities than clump B, with a gradient of 0.1\,kms$^{-1}$\,arcsec$^{-1}$ from the east to the west (corresponding to 25.7\,kms$^{-1}$\,pc$^{-1}$ at 800\,pc). Indeed at low velocities the $^{13}$CO channel maps show a west-south elongated clump where first clump B then clump A becomes more brighter. Between $-$18.41 and $-$17.87\,kms$^{-1}$ several clumps form an S-shaped structure and above $-$17.78\,kms$^{-1}$ the western arc of the structure starts to disintegrate. Around $-$17.35\,kms$^{-1}$ only clump C and extended emission roughly around clump A are present. At high velocities clump A becomes brighter again and another clump appears at 10$\arcsec$ to the northeast at $-$16.29\,kms$^{-1}$. The two clumps are present until around $-$14.9\,kms$^{-1}$.

Emission at the center is present between $-$19.6 and $-$16.7\,kms$^{-1}$ but it is never centered on RNO\,1B. Clump A coincides with one of the sources detected in the 2.7\,mm continuum. The shape of this source suggests that it may be the superposition of smaller objects and one of those might correspond to RNO\,1C. This continuum source roughly coincides with RNO\,1D measured by \citet{weintraub1993} and with VLA\,1 detected by \citet{anglada1994} as well. The brightest continuum source to the northeast coincides with IRAS\,00338+6312 and with a small $^{13}$CO and C$^{18}$O clump. This source was also detected as VLA\,3 by \citet{anglada1994}. VLA\,2 was also detected in the continuum and in $^{13}$CO as well to the southest from RNO\,1B. It also coincides with RNO\,1G \citet{weintraub1993}. The brighter source to the west of it has no counterpart in the CO maps and was not measured in IR or sub-millimeter before. We did not detect RNO\,1F.

\begin{figure*}[tp]
\includegraphics[width=\linewidth, trim=0 1cm 0 0]{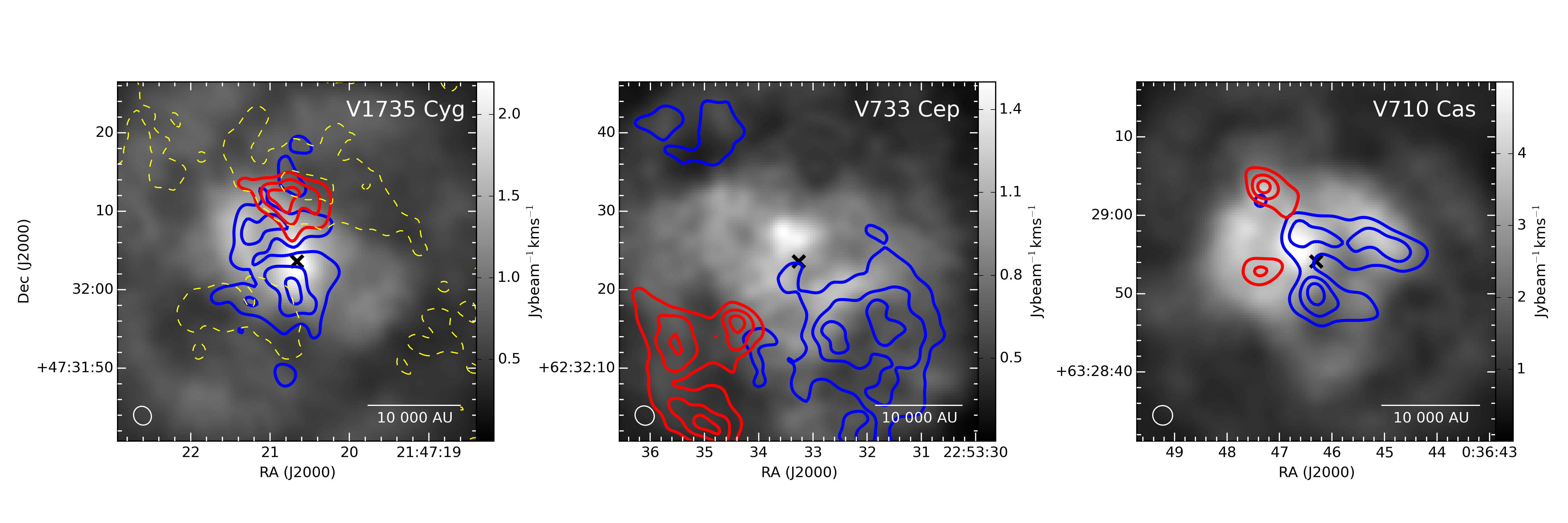}
	\caption{Outflow features around V1735\,Cyg, V733\,Cep and V710\,Cas. The greyscale shows the integrated intensity of the line similarly as on the moment map, black crosses mark the position of the FUors and blue and red lines the integrated intensity contours on the line wings at 50, 70 and 90\% of the peak value. The blue and red line wings were integrated between the following velocities: $-$0.2...2 (blue) and 7...9.4\,kms$^{-1}$ (red) towards V1735\,Cyg; $-$12.9...$-$10.7 (blue) and $-$7.4...$-$6.0\,kms$^{-1}$ (red) towards V733\,Cep; $-$23.3...$-$19.6 (blue), and $-$15.6...$-$13.3\,kms$^{-1}$ (red) towards V710\,Cas. The yellow dashed contours mark the $^{13}$CO FWHM linewidth contours at 4 and 5\,kms$^{-1}$.}
	\label{fuor_outflows}
\end{figure*}

Integrating the $^{13}$CO line wing emission between $-$23.3- $-$19.6, and $-$15.6- $-$13.3\,kms$^{-1}$ (Fig.\ref{fuor_outflows}) the blue lobe extends mainly to the west-southwest and the red lobe consists of two distinct clumps to the east-northeast from the center. The orientation of the lobes mark the same axis as the outflow detected by \citet{evans1994} in $^{12}$CO and by \citet{yang1995} in CS. The same structure can be seen by integrating the C$^{18}$O line wings. However, the emission at these low and high velocities could be just originating in clumps instead of an outflow.

The temperature distribution in the area does not show a clear peak around the center but we see values mostly above 40-50\,K along the arc-shaped structure in the $^{13}$CO map and values of 20-30\,K to the south and on both sides of RNO\,1B. The C$^{18}$O column density map shows higher peaks mainly around and between clump A, IRAS\,00338+6312 and clump B with values above 5\,$\times$\,10$^{16}$\,cm$^{-2}$. Similarly to V1735\,Cyg, calculating the millimeter optical depths shows that except IRAS\,00338+6312 all the other continuum sources are bright due to a temperature effect and not a significant density enhancement. Using the C$^{18}$O emission, clump A with T\,=\,38\,K average temperature has a mass of 1.8\,$M_{\odot}$ and clump B with T\,=\,41\,K has a mass of 0.77\,$M_{\odot}$. 

 \section{Discussion}
 \label{sectdisc}

The CO emission around all our targets shows an extended, several thousands of AU diameter structure, where the intensity of the emission decreases steeper or less steep towards the edge of the primary beam area. Inside this area the distribution of the CO emission is mostly clumpy and complex. There are two cases where the detected $^{13}$CO emission has one significant peak that is centered on the targeted FUors: around V1057\,Cyg with a radius of 3000\,AU and a mass of 0.21\,$M_{\odot}$ and around V1735\,Cyg with a radius of 3400\,AU and a mass of 0.51\,$M_{\odot}$. These clumps are roughly spherical, clearly heated by the central stars, show strong density enhancements and are also detected in the continuum. Similar clumps were detected centered on V1515\,Cyg, V2492\,Cyg and V2493\,Cyg with masses of 0.26, 0.17 and 0.024\,M$_{\odot}$, respectively, but the distribution of the CO emission around these FUors has multiple peaks and shows several clumps. The central clumps of V1515\,Cyg and V2492\,Cyg are heated and these two are detected in the continuum as well. No CO clumps centered on V733\,Cep and V710\,Cas were found but other structures do appear in these areas. In the case of V710\,Cas, clump A is located between the two members of the FUor-system, RNO\,1B and RNO\,1C. This clump is detected also in the continuum. Several other CO clumps in the area also coincide with sub-millimeter, IR or radio objects in previous measurements. 

Considering the match between the detected CO line velocities and the characteristic velocities of the respective star forming regions, the extended circumstellar clouds that roughly fill the primary beam (45.8$\arcsec$ diameter that corresponds to radii of 12600$-$22900\,AU, depending on the distance) around the targeted FUors are clearly associated with the stars \citep{dobashi1994, yonekura1997}. Calculating the $M_{\rm total}$ total mass inside the primary beam using an average temperature we get masses between 1$-$20\,$M_{\odot}$ (Table \ref{mass_vs_si}). Clearly most of the material on this spatial scale will not end up accreting onto the central star, thus these circumstellar structures are not identified as real envelopes. For V1057\,Cyg, V1515\,Cyg, V2492\,Cyg and V1735\,Cyg we find smaller scale clumps coinciding with the position of the star that are heated by the FUors, here we define these as envelopes. In the case of V2493\,Cyg the temperatures are not well constrained and we only detect a small envelope on certain velocities that coincide with the star. In the case of V733\,Cep and V710\,Cas there is no small-scale clump coinciding with the FUors. The presence of many other objects around e.g. V710\,Cas and V2493\,Cyg inside the detected extended circumstellar clouds also reinforces the idea that these are associated clouds but not real envelopes. We also note that the cleaning process of the maps may cause artifacts and noisy areas at the edge of the primary beam in the maps. For the targets where the $^{13}$CO and C$^{18}$O emission is not as centrally peaked (e.g. at V1515\,Cyg, V2493\,Cyg or V733\,Cep) as at V1057\,Cyg or V1735\,Cyg, this might cause some confusion in determining the radius of the extended structure. It could happen that we see only a part of a much larger cloud that is cut off by the primary beam and the cleaning process into a primary beam sized, circular structure. This however is less likely, since we do not see increased noise at the edges, but a clear drop in intensity in all directions. Thus the detected extended clouds are probably real. The central temperature, density, the average temperature and density and the masses of the detected envelopes defined above are summarized in Table \ref{mass_vs_si}.

According to previous $^{12}$CO mapping and optical spectra, all our targets (or neighbouring sources) drive outflows \citep{evans1994, levreault1988}. For three of them the density is enough that the lobes appear even on the $^{13}$CO line wings in our observations. The outflow towards V1735\,Cyg is clearly centered on the star, it is perpendicular to the major axis of the larger scale extended circumstellar cloud and generally oriented the same way as previous, lower resolution molecular studies showed. The line wings towards V710\,Cas could just be blended line components or, considering the large number of IR and radio sources indentified in the region, showing outflow lobes centered on other stars. The situation might be similar towards V733\,Cep where the extended outflow lobes are centered on an unknown source south from the FUor.

Both the continuum-based mass calculation and line-based mass and column density calculation depend strongly on the used envelope temperatures. For most of our target regions the excitation temperatures derived from the optically thick $^{13}$CO emission are consistent with the 20$-$50\,K temperatures used by studies of YSO envelopes \citep{andre1994, hogerheijde1999, sandell2001, lommen2008, dunham2012, green2013, kospal2013}. The $APEX$ study by \citet{kospal2017} derived lower temperatures (5$-$7\,K) for lower mass envelopes and higher temperatures (>\,50\,K) for high-mass envelopes. If we do not consider V2493\,Cyg, V1735\,Cyg with the highest envelope mass shows the highest average temperature (25\,K) but the lower-mass envelope of V2492\,Cyg is around this value as well. Somewhat higher temperatures are measured around the binary FUor, RNO\,1B/C (30$-$50\,K). One explanation for such high values could be the high energy input from several strong outflows driven by young sources in the region.

The different evolutionary stages of YSOs can be distinguished by investigating their SED in the sub-millimeter and IR, which is also connected to the amount of circumstellar material associated with the central star. According to the dust continuum studies of YSOs in $\rho$\,Ophiuchi and Taurus by \citet{andre1994}, \citet{bontemps1996} and \citet{moriarty1994}, envelope masses higher than 0.5\,$M_{\odot}$ are characteristic for Class\,0 protostars that are still in their main accretion phase. Class\,I objects were found with masses between 0.015 and 0.15\,$M_{\odot}$ with a median of 0.06\,$M_{\odot}$ and Class\,II objects have masses of 0.002$-$0.15\,$M_{\odot}$ with a median of 0.01\,$M_{\odot}$. While the majority of Class\,III sources were undetected at 1.3\,mm, the circumstellar material around Class\,II stars remained usually unresolved with the approximately 1000\,AU (12$\arcsec$) resolution of the IRAM 30\,m telescope. Additionally, Class\,I objects were detected mostly as extended, resolved sources with radii of less than a few thousand AU. Apart from the dust emission from the disk, part of the emission from the envelope is also included in the continuum detections of Class\,I sources \citep{terebey1993}.

\begin{table*}[tp]
	\centering
    \footnotesize
		\caption{Envelope masses and the silicate feature at out targets}
       % \resizebox{.5\textwidth}{!}{
		\begin{tabular}{l l c l c c l l c}
		\hline
		\multicolumn{1}{c}{Target} & \multicolumn{1}{c}{T$_{\rm c}$} & \multicolumn{1}{c}{$N_{\rm c}$}  &  \multicolumn{1}{c}{T$_{\rm env}$} & \multicolumn{1}{c}{$N_{\rm env}$}  & \multicolumn{1}{c}{$R_{\rm env}$} &  \multicolumn{1}{c}{$M_{\rm env}$} & \multicolumn{1}{c}{$M_{\rm total}$} & Si band  \\
		& [K]  & [10$^{16}$\,cm$^{-2}$] & [K]  & [10$^{16}$\,cm$^{-2}$] &  [AU] & \multicolumn{1}{c}{[$M_{\odot}$]} & \multicolumn{1}{c}{[$M_{\odot}$]} & \multicolumn{1}{c}{\,} \\
		\hline 
		\hline
V1057\,Cyg  & 17 & 0.81 & 16.4 & 0.84 & 3000 & 0.21 & 1.25 & em \\
V1515\,Cyg & 14.2 & 0.25 & 11 & 0.29 & 5000 & 0.26 & 3.1 & em \\
V2492\,Cyg & 24.1 & 1.68 & 23.5 & 1.10 & 1925 & 0.17 & 1.9 & ... \\
V2493\,Cyg & ... & ... & 37 & 0.29 & 1650  & 0.024 & 3.15 & ... \\
V1735\,Cyg & 24.5 & 0.41 & 19.7 & 0.51 & 3400 & 0.51 & 6.9 & abs \\
V733\,Cep & 14 & 0.12 & ... & ... & ... & \multicolumn{1}{c}{...} & 4.5 & abs \\
V710\,Cas & 31.9 &  3.62 & ... & ... & ... & \multicolumn{1}{c}{...} & 19.6 & abs \\
\hline
		\end{tabular}%}
    \tablefoot{The columns are: (1) target name; (2) temperature at the target position; (3) C$^{18}$O column density at the target position; (4) average temperature of the central clump (envelope); (5) average C$^{18}$O density of the central clump; (6) radius of the detected central clump; (7) the mass of the central clump; (8) total mass measured inside the primary beam (45.8$\arcsec$); (4) appearance of the silicate feature \citep{green2006, larsson2000}.}
	\label{mass_vs_si}
\end{table*}

The line-based masses of the detected FUor envelopes presented in this paper span two orders of magnitude, from 0.024 to 0.51\,$M_{\odot}$. Similar masses were derived by \citet{kospal2017} using $^{12}$CO and $^{13}$CO measurements by $APEX$ for eight southern and equatorial FUors. The continuum-based masses of the three strongly detected FUors (V1057\,Cyg, V2492\,Cyg, V1735\,Cyg) are fairly uniform, around 0.3$-$0.4\,$M_{\odot}$. Some correlation of the dust masses with the line-based values can be seen for the three targets with no significant millimeter detections (V2493\,Cyg, V733\,Cep, V710\,Cas), since they have also no or very small CO envelopes. The two targets with similar line-based masses (V1057\,Cyg and V2492\,Cyg) also have very similar continuum masses. However, V1735\,Cyg with a high line-based envelope mass also shows a continuum mass similar to those two and V1515\,Cyg, that was detected as a weak millimeter source, has the second highest line-based envelope mass. The differences in gas/dust can be explained by the difference in the mass of their circumstellar disk, the differing viewing angle of the system, different CO depletion of grain growth.

Considering the derived envelope masses, the evolutionary states of our FUor targets span both very young and embedded, early Class\,I stages (V1735\,Cyg with a high mass and extended gas envelope) and a more evolved, Class\,II stage (V2493\,Cyg with no continuum detection and a smaller, low-mass envelope). The non-detection of gas features centered on V733\,Cep and V710\,Cas also suggests low-mass envelopes and they are weakly or not detected in the continuum, implying that they are also more evolved, similar to Class\,II objects. The envelope masses of V1057\,Cyg, V1515\,Cyg and V2492\,Cyg implies a Class\,I classification but V1515\,Cyg has a much more extended gas envelope and is only weakly detected at 2.7\,mm.

\citet{gramajo2014} used the model set of \citet{whitney2003a, whitney2003b} and \citet{robitaille2006} to model the SED of almost all known FUors and derived the parameters of the stars, the disks and the envelopes. They determined that in general FUors have higher mass disks and higher disk accretion rates than quiescent Class\,I and Class\,II sources in Taurus. Their best fitting models of our targets have outer envelope radii of a few thousands of AU, usually two times the values we calculated. For V2492\,Cyg both the envelope radii (2000\,AU) and the envelope mass (0.2\,$M_{\odot}$) from their model are very close to our line-based values. Amongst our targets they find the lowest envelope mass for V2493\,Cyg (0.06\,$M_{\odot}$) and the highest for V1735\,Cyg (0.9\,$M_{\odot}$) which agrees with our results as well. However, they derive almost a magnitude smaller masses for the envelopes of V1057\,Cyg and V1515\,Cyg. There is no clear correlation between their derived disk masses and the detectability of our targets in the 2.7\,mm continuum. They classified V1515\,Cyg, RNO\,1B/C and V1735\,Cyg as Class II sources and V1057\,Cyg, V2492\,Cyg and V2493\,Cyg as Class I sources. They performed no modelling for V733\,Cep.

Based on the appearance of the 10\,$\mu$m silicate feature \citet{quanz2007b} defined two categories specifically amongst FUors: Category\,1 stars show the feature in absorption, which is explained with a dusty and icy envelope surrounding the object. Category\,2 stars show the feature in emission that arises from the surface layer of the circumstellar accretion disk. There is also evidence of grain growth in the emission profiles of Category 2 stars. These two groups can be explained if we assume that Category\,1 FUors are younger, more similar to embedded Class\,0 or Class\,I objects, while Category\,2 FUors are more evolved. The accretion disk is visible to the observer because much of the envelope has already dispersed or accreted onto the central star and a wide cavity has opened up. The explanation is in agreement with theoretical model predictions that expect FUor outbursts to be recurring and then, when the envelope is cleared up, the star to enter a more quiescent phase. This way FUors can be the link between Class\,I and Class\,II stars and FUor outbursts may play the key role in this transition. 

From this paradigm it follows that Category\,2 FUors must have less massive envelopes than Category\,1 FUors. According to \textit{Spitzer} and \textit{ISO} data \citep{green2006, larsson2000} the spectra of V710\,Cas, V1735\,Cyg and V733\,Cep show the silicate feature in absorption, V1515\,Cyg and V1057\,Cyg show it in emission, while there is no good spectral resolution data of the 10\,$\mu$m band towards V2492\,Cyg and V2493\,Cyg. V1735\,Cyg shows the feature in absorption and this star has the most massive envelope in our sample, while V1057\,Cyg and V1515\,Cyg have lower mass envelopes and show the silicate feature in emission. This appears to confirm the correlation between the two parameters, however, the two other Category\,1 FUors (V733\,Cep and V10 Cas) have no detected CO clumps associated with them (Table \ref{mass_vs_si}), thus seem to be more evolved systems. The results are also ambiguous since two FUors with low-mass envelopes have no 10\,$\mu$m spectra available. Correlating the two parameters this way might be challenging since while silicate emission or absorption originates from the disk (the inner few AU for a low mass star) and is spatially unresolved, CO structures can be found on both smaller and larger scales around the objects. For individual objects the viewing angle of the equatorial plane of the object also has a large effect on the silicate spectral shape. Even if the envelope is already depleted and the object is more evolved, viewing it edge-on will result in the absorption of the silicate band. We note however that since our objects are generally visible in the optical regime it is probable that even if they are embedded we see them somewhat from above the disk plane, around the direction of their envelope cavity. 

Interestingly there is also a correlation of $M_{\rm total}$ with the 10\,$\mu$m spectral: the stars showing the silicate feature in emission have total masses of 1.25 and 3.1\,$M_{\odot}$ and those that show it in absorption have total masses above 4.5\,$M_{\odot}$. However, calculating the total mass inside the primary beam clearly includes many other sources and their potential envelopes as well (as it is the case with V710\,Cas). This way $M_{\rm total}$ provides information about the larger scale environment of the star which still has an effect on the detected spectral features. In the case of V733\,Cep and V710\,Cas, where the non-detection in the continuum and the lack of a central CO clump suggests that their envelope has already strongly depleted, the absorption feature might be created by the large scale structures around them. In the case of V1057\,Cyg and perhaps V1515\,Cyg, which seem like embedded, Class\,I sources, we might see the 10\,$\mu$m emission since we observe the stars from the direction of their circumstellar cavity.

\section{Summary and Conclusions}

We presented an interferometric $^{13}$CO and C$^{18}$O survey of northern FU\,Orionis-type stars. Strong $^{13}$CO and C$^{18}$O emission was detected around each of our targets in a few thousand AU spatial scale, tracing extended circumstellar clouds that show sub-structure. The smaller scale distribution of the emission peaks at or close to the optical positions of the FUors and is generally composed of a few significant or many clumps with complicated kinematic structure. In most cases the $^{13}$CO emission is optically thick and can be used to derive temperatures, while C$^{18}$O is optically thin and so suitable to determine the densities and masses of the detected structures. The calculated temperatures are mostly in the expected range for envelopes around low-mass stars (20$-$50\,K). The peak C$^{18}$O column densities in the envelopes range from a few times 10$^{15}$ to 10$^{16}$\,cm$^{-2}$ which gives H$_2$ column densities of 0.5$-$5\,$\times$\,10$^{22}$. The temperature distribution also implies the heating of the detected small-scale central clumps by the targeted FUors.

\textit{V1057\,Cyg}: We found evidence of a rotating spherical clump centered on the star with a mass of 0.21\,$M_{\odot}$, that is also detected in the 2.7\,mm continuum, implying a fairly large and dense envelope. The star was classified as a Class\,II object by its flat SED \citep{green2013} but its envelope mass and size suggests a younger, Class\,I object. Viewing the source from the direction of a large polar cavity that was previously proposed by \citet{green2006} could explain the silicate emission feature and the SED.

\textit{V1515\,Cyg}: We derive a similar mass for the envelope as for V1057\,Cyg, and the silicate feature also appears in emission, which again, could be explained by viewing the source from the direction of its polar hole \citep{goodrich1987}. The SED is flat, suggesting a Class\,II object, and the continuum mass of the source is very low, which might also imply a somewhat more evolved state or an especially low-mass circumstellar disk. We find the lowest envelope temperature here in our sample. A roughly circular, ring-like, clumpy structure was detected around the FUor on the $^{13}$CO channel maps that might trace the walls of an expanding shell, with a radius of $\approx$\,5000\,AU. 

\textit{V2492\,Cyg}: The envelope mass is very similar to V1057\,Cyg and it is also detected at 2.7\,mm, suggesting a younger, embedded, Class\,I object. A large-scale velocity gradient was observed around the FUor, that may originate from the asymmetric compression of the material in the elongated, trunk-like cloud by the ionization front generated by close-by luminous stars. We found many clumps behind the ionization front, differing in density and mass, but many of them are clearly heated. 

\textit{V2493\,Cyg}: An extended, clumpy cloud structure with high temperatures was observed around the target. We measured the lowest envelope mass and the smallest central CO clump here in our sample, the derived mass is close to the median envelope mass of Class\,II stars. The non-detection of the source in $SMA$ 1.3\,mm \citep{dunham2012} and in our 2.7\,mm maps also suggests that it is more evolved.

\textit{V1735\,Cyg}: While the SED implies a Class\,II object, we found a massive, warm envelope that is detected at 2.7\,mm as well, more characteristic of younger Class\,I stars. Perpendicular to a flattened, elliptical extended cloud structure, clear traces of a $^{13}$CO outflow was found. The silicate feature appears in absorption which also implies an earlier evolutionary state.

\textit{V733\,Cep}: The environment around V733\,Cep consists of two clumps on an offset from the position of the star. We found no CO or continuum emission that is clearly centered on the star, implying a later evolutionary state, but the silicate feature is seen in absorption. This might suggests the effect of the larger-scale circumstellar material on the spectral features. The FUor or another undetected object might also drive an outflow in the region.

\textit{RNO\,1B/C}: The envelope around the target was not detected but many clumps around the binary FUor system were. Some of them appear in the continuum as well, while the target was not detected at millimeter wavelength. The brightest detected CO clump with a mass of 1.8\,$M_{\odot}$ may be a circumbinary structure around RNO\,1B and RNO\,1C, and the reason behind the observed silicate absorption feature. 

The classification of our objects into the FUor categories established by MIR observations of the silicate feature faces many problems. The determination of the exact evolutionary state is complicated by the unknown viewing angle of the sources, other embedded or young objects in the region and the lack of even higher resolution spectral mapping around. However, we find that some of our targets exhibit parameters similar to very young, embedded Class\,I stars and some of them seem more evolved with only small envelopes, resembling early or late Class\,II YSOs. This diversity seems to reinforce the theory that FUors are Class\,I and Class\,II young stellar sources in a special evolutionary phase with bright eruptions and that these infall-driven FUor eruptions are the main driving force of this transition.

\begin{acknowledgements}

This research was supported by the Momentum grant of the MTA CSFK Lend\"ulet Disk Research Group. This work is based on observations carried out under project number VA6C with the IRAM Plateau de Bure Interferometer and under project number 260-11 with the IRAM 30 m Telescope. IRAM is supported by INSU/CNRS (France), MPG (Germany) and IGN (Spain). We are grateful to Michael Bremer for his help with the IRAM data reduction. The authors also wish to thank the anonymous referee for their valuable comments and suggestions.

\end{acknowledgements}

\bibliography{FUor_CO_v2}
\bibliographystyle{aa}

\begin{appendix}

\section{$^{13}$CO and C$^{18}$O channel maps}

For the channel maps see the Appendix of the online article at http://www.aanda.org.

\end{appendix}


\end{document}